\documentclass[11pt]{article}

\usepackage{cite}
\usepackage{subfigure}
\usepackage{multirow}
\usepackage{helvet}
\usepackage{amsmath}
\usepackage{amssymb}
\usepackage{setspace}
\usepackage{setspace}
\usepackage{graphicx}
\usepackage{epsfig}
\usepackage{empheq}
\usepackage{subfigure}

\usepackage[a4paper,top=3cm,bottom=3cm,left=2.5cm,right=2.5cm,bindingoffset=0mm]{geometry}

\def\be{\begin{equation}}
\def\ee{\end{equation}}

\begin{document}

\titlepage               
 \begin{flushright}                                                    
 IPPP/18/90                                                                                                     
  \end{flushright}                                      

 \vspace*{0.5cm}
\begin{center}                                                    
{\Large \bf Exclusive LHC physics with heavy ions: SuperChic 3}\\

\vspace*{1cm}
                                                   
L.A. Harland--Lang$^{1}$, V.A. Khoze$^{2,3}$, M.G. Ryskin$^{3}$ \\                                                
                                                   
\vspace*{0.5cm}
${}^1$Rudolf Peierls Centre for Theoretical Physics, University of Oxford,
Clarendon Laboratory, Parks Road, Oxford OX1 3PU, United Kingdom. \\                                                                                      
${}^2$Institute for Particle Physics Phenomenology, University of Durham, Durham, DH1 3LE \\
${}^3$Petersburg Nuclear Physics Institute, NRC Kurchatov Institute, Gatchina, \linebreak[4]St. Petersburg, 188300, Russia    
  
\vspace*{1cm}                         

\begin{abstract}

\noindent  We present results of the updated \texttt{SuperChic 3} Monte Carlo event generator for central exclusive production. This extends the previous treatment of proton--proton collisions to include heavy ion (pA and AA) beams, for both photon and QCD--initiated production, the first time such a unified treatment of exclusive processes has been presented in a single generator. To achieve this we have developed a theory of the gap survival factor in heavy ion collisions, which allows us to derive some straightforward results about the $A$ scaling of the corresponding cross sections. We compare against the recent ATLAS and CMS measurements of light--by--light scattering at the LHC, in lead--lead collisions. We find that the background from QCD--initiated production is expected to be very small, in contrast to some earlier estimates. We also present results from new photon--initiated processes that can now be generated, namely the production of axion--like particles, monopole pairs and monopolium, top quark pair production, and the inclusion of $W$ loops in light--by--light scattering.

\end{abstract}
                                   
\end{center}  

\section{Introduction}

Central Exclusive Production (CEP) is the reaction

\be
hh \to h\,+\, X\,+\,h
\ee
where `+' signs are used to denote the presence of large rapidity gaps, separating the system $X$ from the intact outgoing hadrons $h$. This simple signal is associated with a broad and varied phenomenology, from low energy QCD to high energy BSM physics, see~\cite{Albrow:2008pn,Albrow:2010yb,Tasevsky:2014cpa,Harland-Lang:2014lxa,Harland-Lang:2014dta,N.Cartiglia:2015gve} for reviews. Consequently an extensive experimental programme is planned and ongoing at the LHC, with dedicated proton tagging detectors installed and collecting data in association with both ATLAS and CMS~\cite{AFP1,Albrow:1753795}, while multiple measurements using rapidity gap vetoes have been made by LHCb and ALICE.

\begin{figure}
\begin{center}
\includegraphics[scale=0.45]{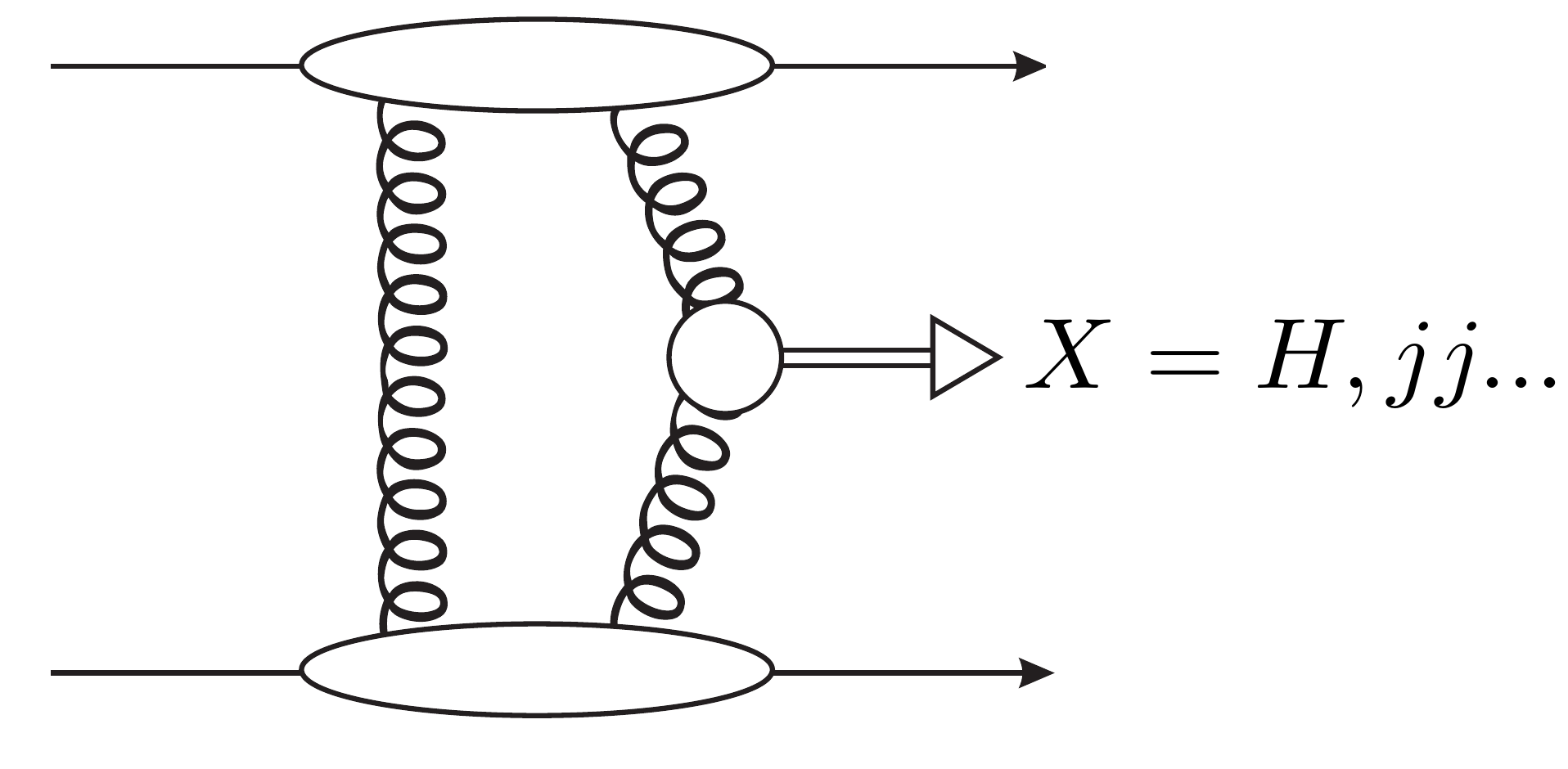}
\includegraphics[scale=0.45]{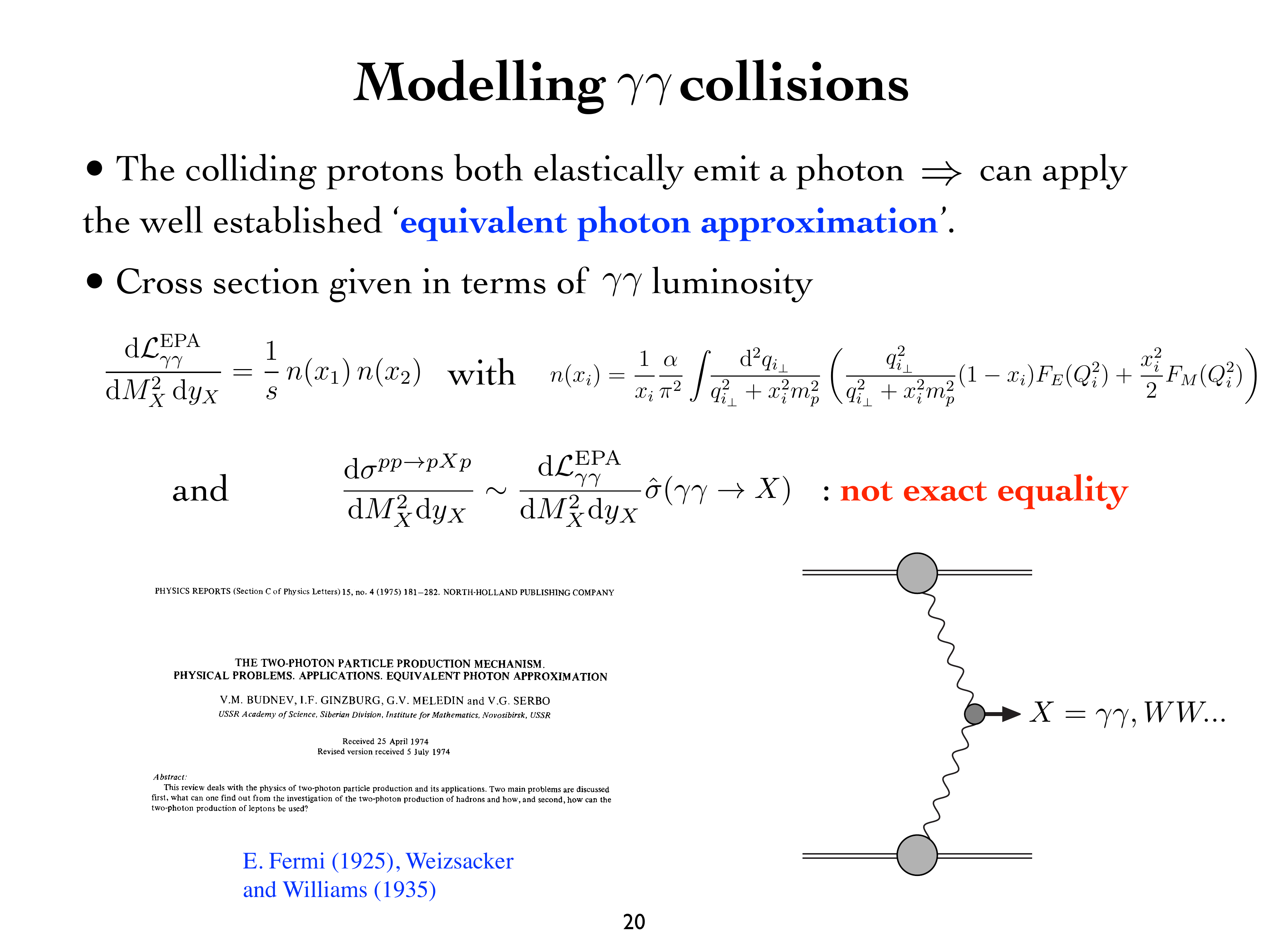}
\caption{Schematic diagrams for (left) QCD and (right) photon initiated CEP.}
\label{fig:pCp}
\end{center}
\end{figure}

CEP  may proceed via either QCD or photon--induced interactions, see Fig.~\ref{fig:pCp}, as well as through a combination of both, namely via photoproduction. Although producing the same basic exclusive signal, each mechanism is distinct in terms of the theoretical framework underpinning it and the phenomenology resulting from it. The QCD--initiated mechanism benefits from a `$J^{PC}=0^{++}$' selection rule, permitting the production of a range of strongly interacting states in a precisely defined gluon--rich environment, while also providing a non--trivial test of QCD in a distinct regime from standard inclusive production. The framework for describing photon--initiated production is under very good theoretical control, such that one can in effect use the LHC as a photon--photon collider; this well understood QED initial state provides unique sensitivity to beyond the Standard Model (BSM) effects. Photoproduction can for example provide a probe of  low $x$ QCD effects such as gluon saturation in both proton and nuclear targets. For further information and reviews, see~\cite{Khoze:2001xm,Albrow:2010yb,Tasevsky:2014cpa,Harland-Lang:2014lxa,Harland-Lang:2014dta}.

As mentioned above, a range of  measurements have been made and are ongoing at the LHC. To support this experimental programme, it is essential to provide Monte Carlo (MC) tools to connect the theoretical predictions for CEP with the experimental measurements.  For this reason the authors have previously produced the publicly available \texttt{SuperChic} MC~\cite{HarlandLang:2009qe,HarlandLang:2010ep}, subsequently upgraded to version 2 in~\cite{Harland-Lang:2015cta}. This generates a wide range of QCD and photon--initiated processes in $pp$ collisions, with the former calculated using the perturbative `Durham' approach. In addition, this includes a fully differential treatment of the soft survival factor, that is the probability of no additional soft particle production, which would spoil the exclusivity of the event. 

Other available MC implementations include: FPMC~\cite{Boonekamp:2011ky}, which generates a smaller selection of final--states and does not include a differential treatment of survival effects, although it also generates more inclusive diffractive processes, beyond pure CEP; an implementation of CEP in \texttt{Pythia} described in~\cite{Lonnblad:2016hun}, which provides a full treatment of initial--state showering effects for a small selection of processes, allowing both pure CEP and semi--exclusive production to be treated on the same footing, while the survival factor is included via the standard \texttt{Pythia} treatment of multi-particle interactions (MPI); the \texttt{Starlight} MC~\cite{Klein:2016yzr} generates a range of photon--initiated and photoproduction processes in heavy ion collisions; \texttt{ExHuME}~\cite{Monk:2005ji}, for QCD--initiated production of a small selection of processes; \texttt{CepGen}~\cite{Forthomme:2018ecc}, which considers photon--initiated production but aims to allow the user to add in arbitrary processes; for lower mass QCD--initiated production, the \texttt{Dime}~\cite{Harland-Lang:2013dia}, \texttt{ExDiff}~\cite{Ryutin:2017qii} and \texttt{GenEx}~\cite{Kycia:2017ota} MCs.

As discussed above,  the \texttt{SuperChic} MC aims to provide a treatment of all mechanisms for CEP, both QCD and photon initiated, within a unified framework. However, so far it has only considered the case of proton--proton (or proton--antiproton) collisions; CEP with heavy ion (pA and AA) beams, so--called `ultra--peripheral' collisions (UPCs), have not been included at all. Such processes are of much interest, with in particular the large photon flux $\sim Z^2$ per ion enhancing the signal for various photon--initiated processes. In this paper we therefore extend the MC framework to include both proton--ion and ion--ion collisions, for arbitrary beams and in both QCD and photon--initiated production. 

Indeed, a particularly topical example of this is the case of light--by--light (LbyL) scattering, $\gamma\gamma \to \gamma\gamma$, evidence for which was found by ATLAS~\cite{Aaboud:2017bwk} and more recently CMS~\cite{dEnterria:2018uly}. These represent the first direct observations of this process, and these data already show sensitivity to various BSM scenarios~\cite{Knapen:2016moh,Ellis:2017edi}. However, one so--far unresolved question is the size of the potential background from QCD--initiated production, $gg \to \gamma\gamma$, which in both analyses was simply taken from the \texttt{SuperChic} prediction in $pp$ collisions and scaled by $A^2 R^4$, where the factor $R\sim 0.7$ accounted for gluon shadowing effects, that is assuming that all $A$ nucleons in each ion can undergo CEP. While the normalization of this baseline prediction was in fact left free and set by data--driven methods, it is nonetheless important to address whether such a prediction is indeed reliable, by performing for the first time a full calculation of QCD--initiated production in heavy ion collisions. We achieve this here, and as we will see, predict that this background is much lower than previously anticipated.

A further topical CEP application is the case of high mass production of electroweakly coupled BSM states, for which photon--initiated production will be dominant at sufficiently high mass~\cite{Khoze:2001xm}. Events may be selected with tagged protons in association with central production observed by ATLAS and CMS, during nominal LHC running. There are possibilities, for example, to probe anomalous gauge couplings (see~\cite{Baldenegro:2017aen} and references therein) and search for high mass pseudoscalar states~\cite{Baldenegro:2018hng} in these channels, accessing regions of parameters space that are difficult or impossible to reach using standard inclusive methods. With this in mind, we also present various updates to the photon--initiated production channels. Namely, we provide a refined calculated of Standard Model (SM) LbyL scattering, including the $W$ loops that are particularly important at high mass, as well as generating axion--like particle (ALP), monopole pair and monopolium production. We also include photon--initiated top quark pair production. We label the MC including these updates \texttt{SuperChic 3}.

The outline of this paper is as follows. In Section~\ref{sec:heavyion} we present details of the implementation of CEP in pA and AA collisions, for both photon and QCD--initiated cases. In Section~\ref{sec:newproc} we discuss the new photon--initiated processes that are included in the MC. In Section~\ref{sec:LbyL} we take a closer look at LbyL scattering, comparing in detail to the ATLAS and CMS data, and considering both the photon--initiated signal and QCD-initiated background. In Section~\ref{sec:superchic} we summarise the processes generated by \texttt{SuperChic 3} and provide information on its availability. In Section~\ref{sec:conc} we conclude, and in Appendix~\ref{ap:incoh} we present some analytic estimates of the expected scaling with $A$ of the QCD--initiated production process in pA and AA collisions, supporting our numerical findings.

\section{Heavy Ion Collisions}\label{sec:heavyion}

We first consider the photon--initiated production, before moving on to consider the QCD--initiated case.

\subsection{$\gamma\gamma$ collisions -- unscreened case}\label{sec:gamunsc}

\begin{figure}
\begin{center}
\includegraphics[scale=0.6]{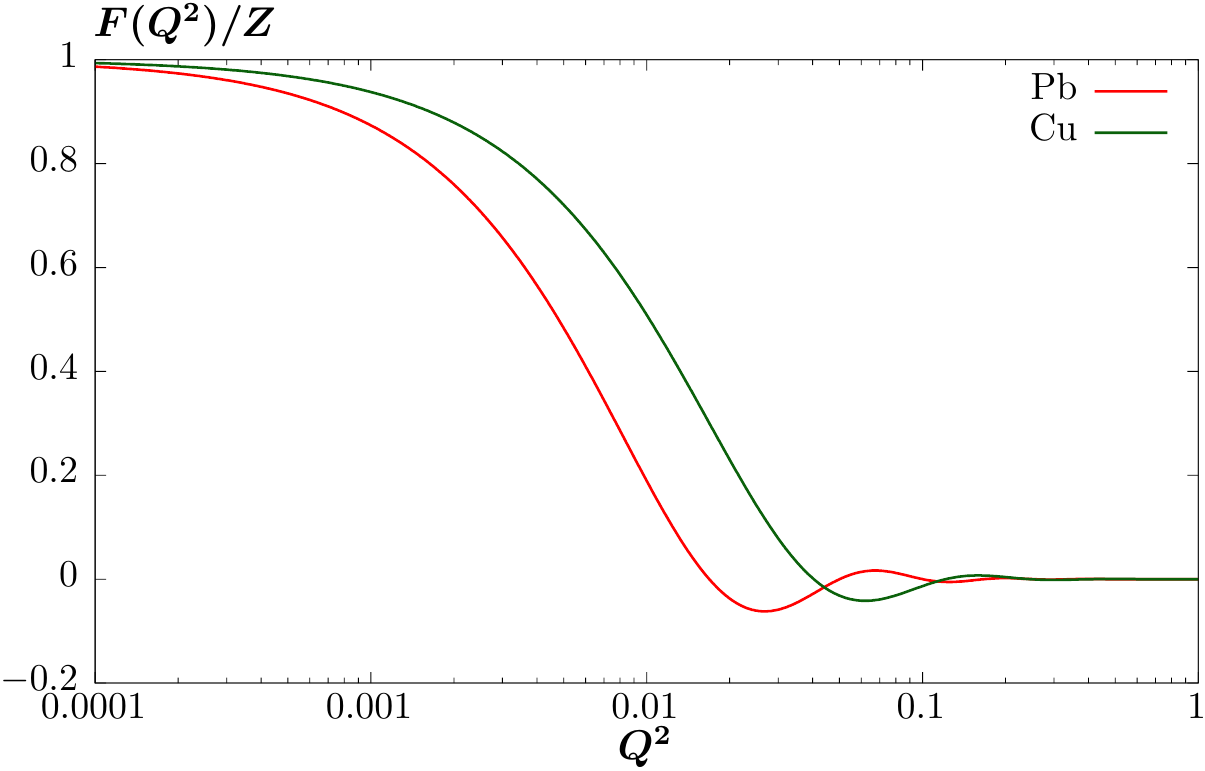}
\caption{Normalized charge form factor due to lead and copper ions.}
\label{fig:ff}
\end{center}
\end{figure}

For photon--initiated production in heavy ion collisions, ignoring for now the possibility of additional ion--ion interactions, we can apply the usual equivalent photon approximation~\cite{Budnev:1974de}. The cross section for the production of a system of mass $M_X$ and rapidity $Y_X$ is given by
\begin{align}
\sigma_{N_1 N_2 \to N_1 X N_2} &= \int {\rm d} x_1 {\rm d}x_2 \,n(x_1) n(x_2)\hat{\sigma}_{\gamma \gamma \to X}\;,\\
&= \int {\rm d}M_X {\rm d}Y_X \frac{2 M_X}{s}  \,n(x_1) n(x_2) \hat{\sigma}_{\gamma \gamma \to X}\;,
\end{align}
where the photon flux is 
\be
n(x_i)=\frac{\alpha}{\pi^2 x_i}\int \frac{{\rm d}^2q_{i_\perp} }{q_{i_\perp}^2+x_i^2 m_{N_i}^2}\left(\frac{q_{i_\perp}^2}{q_{i_\perp}^2+x_i^2 m_{N_i}^2}(1-x_i)F_E(Q_i^2)+\frac{x_i^2}{2}F_M(Q_i^2)\right)\;,
\ee
in terms of the transverse momentum $q_{i\perp}$ and longitudinal momentum fraction $x_i$ of the parent nucleus carried by the photon\footnote{Correspondingly, we have $s= A_1 A_2 s_{nn}$, where $s_{nn}$ is the squared c.m.s. energy per nucleon and $A_i$ is the ion mass number.}. The modulus of the photon virtuality, $Q^2_i$, is given by
\begin{equation}\label{eq:qi}
Q^2_i=\frac{q_{i_\perp}^2+x_i^2 m_{N_i}^2}{1-x_i}\;,
\end{equation}
For the proton, we have $m_{N_i}=m_p$ and the form factors are given by
\begin{equation}
F_M(Q^2_i)=G_M^2(Q^2_i)\qquad F_E(Q^2_i)=\frac{4m_p^2 G_E^2(Q_i^2)+Q^2_i G_M^2(Q_i^2)}{4m_p^2+Q^2_i}\;,
\end{equation}
with
\begin{equation}
G_E^2(Q_i^2)=\frac{G_M^2(Q_i^2)}{7.78}=\frac{1}{\left(1+Q^2_i/0.71 {\rm GeV}^2\right)^4}\;,
\end{equation}
in the dipole approximation, where $G_E$ and $G_M$ are the `Sachs' form factors. For the heavy ion case the magnetic form factor is only enhanced by $Z$, and so can be safely dropped. We then have 
\begin{equation}\label{eq:wwion}
F_M(Q^2_i)=0\qquad F_E(Q^2_i)=F_p^2(Q_i^2)G_E^2(Q_i^2)\;,
\end{equation}
where $F_p(Q^2)^2$ is the squared charge form factor of the ion.
Here, we have factored off the $G_E^2$ term, due to the form factor of the protons within the ion; numerically this has a negligible impact, as the ion form factor falls much more steeply, however we include this for completeness. The ion form factor is given in terms of the proton density in the ion, $\rho_p(r)$, which is well described by the Woods--Saxon distribution~\cite{Woods:1954zz}
\be\label{eq:rhop}
\rho_p(r)= \frac{\rho_0}{1+\exp{\left[(r-R)/d\right]}}\;,
\ee
where the skin thickness $d \sim 0.5-0.6$ fm, depending on the ion, and the radius $R \sim A^{1/3}$. The density
 $\rho_0$ is set by requiring that
\be
\int {\rm d}^3  r\,\rho_p(r) = Z\;.
\ee
The total nucleon density $\rho_A$ can be defined in a similar way, and is normalised to the mass number $A$.
The charge form factor is then simply given by the Fourier transform
\be
F_p(|\vec{q}|) = \int {\rm d}^3r \, e^{i \vec{q}\cdot \vec{r}}  \rho_p(r)\;,
\ee
in the rest frame of the ion; in this case we have $\vec{q}^2 = Q^2$, so that written covariantly this corresponds to the $F(Q^2)$ which appears in \eqref{eq:wwion}. In impact parameter space, the coherent amplitude is given by a convolution of the transverse proton density within the ion, and the amplitude for photon emission from individual protons; hence in transverse momentum space we simply multiply by the corresponding form factor. This is shown in Fig.~\ref{fig:ff} for the case of ${}^{63}{\rm Cu}$ and ${}^{208}{\rm Pb}$, for which we take~\cite{Chamon:2002mx}
\be\label{eq:rdep}
R  = (1.31 A^{1/3} - 0.84) \, {\rm fm}\;, \qquad d=0.55 \, {\rm fm}\;,
\ee
for concreteness. The sharp fall off with $Q^2$ is clear, with the form factors falling to roughly zero by $\sqrt{Q^2} \sim 3/R\sim 0.1$ GeV; for the smaller Cu ion this extends to somewhat larger $Q^2$ values.

The above results, which are written at the cross section level, completely define the situation in the absence of screening corrections. However for the purpose of future discussion we can also write this in terms of the amplitude
\be\label{eq:tq1q2}
T(q_{1\perp},q_{2\perp}) = \mathcal{N}_1 \mathcal{N}_2 \,q_{1\perp}^\mu q_{2\perp}^\nu V_{\mu\nu}\;,
\ee  
where $V_{\mu\nu}$ is the $\gamma\gamma \to X$ vertex, and the normalization factors are given by
\be
\mathcal{N}_i = \left(\frac{\alpha}{\pi x_i}(1-x_i)\right)^{1/2}\frac{F(Q_i^2)}{q_{i_\perp}^2+x_i^2 m_{N_i}^2}\;.
\ee
Indeed, the derivation of the equivalent photon approximation at the amplitude level has precisely this Lorentz structure\footnote{Strictly speaking this is only true for the contribution proportional to the electric form factors, see~\cite{Harland-Lang:2015cta} for further discussion; however here we indeed take $F_M=0$.}. This then reduces to the usual cross section level result after noting that we can write 
\begin{align}
q_{1_\perp}^i q_{2_\perp}^j V_{ij} =\begin{cases} &-\frac{1}{2} ({\bf q}_{1_\perp}\cdot {\bf q}_{2_\perp})(\mathcal{M}_{++}+\mathcal{M}_{--})\;\;(J^P_z=0^+)\\ 
&-\frac{i}{2} |({\bf q}_{1_\perp}\times {\bf q}_{2_\perp})|(\mathcal{M}_{++}-\mathcal{M}_{--})\;\;(J^P_z=0^-)\\ 
&+\frac{1}{2}((q_{1_\perp}^x q_{2_\perp}^x-q_{1_\perp}^y q_{2_\perp}^y)+i(q_{1_\perp}^x q_{2_\perp}^y+q_{1_\perp}^y q_{2_\perp}^x))\mathcal{M}_{-+}\;\;(J^P_z=+2^+)\\ 
&+\frac{1}{2}((q_{1_\perp}^x q_{2_\perp}^x-q_{1_\perp}^y q_{2_\perp}^y)-i(q_{1_\perp}^x q_{2_\perp}^y+q_{1_\perp}^y q_{2_\perp}^x))\mathcal{M}_{+-}\;\;(J^P_z=-2^+)
\end{cases}\label{Agen}
\end{align}
where $\mathcal{M}_{\pm \pm}$ corresponds to the $\gamma(\pm) \gamma(\pm) \to X$ helicity amplitude. We then have
\be
\int {\rm d}^2 q_{1\perp}{\rm d}^2 q_{2\perp} |T(q_{1\perp},q_{2\perp}) |^2 = n(x_1)n(x_2) \frac{1}{4} \sum_{\lambda_1 \lambda_2} |\mathcal{M}_{\lambda_1 \lambda_2}|^2\;,
\ee
after performing the azimuthal angular integration on the left hand side.

The cross section is then given by
\be\label{eq:csn}
\sigma_{N_1 N_2 \to N_1 X N_2} =\int {\rm d} x_1 {\rm d}x_2 {\rm d}^2 q_{1\perp}{\rm d}^2 q_{2\perp} \mathcal{PS}_i |T(q_{1\perp},q_{2\perp}) |^2\;,
\ee
where $\mathcal{PS}_i$ is defined for the $2\to i$ process to reproduce the corresponding cross section $\hat{\sigma}$, i.e. explicitly
\be
\mathcal{PS}_1 = \frac{\pi}{M_X^2}\delta(\hat{s}-M^2)\;,\qquad \mathcal{PS}_2 = \frac{1}{64\pi^2 M_X^2} \int {\rm d}\Omega\;.
\ee
It is then straightforward to see that this reduces to the usual equivalent photon result. However, as we will see below, we must work at the amplitude level to give a proper account of screening corrections.

\subsection{$\gamma\gamma$ collisions -- screened case}\label{sec:gamsc}

The inclusion of screening corrections follows in essentially straightforward analogy to the $pp$ case considered in e.g.~\cite{Gotsman:2014pwa,Harland-Lang:2015cta,Khoze:2017sdd}. This is most easily discussed in impact parameter space, for which the average eikonal survival factor is given by
\begin{equation}\label{eq:S2}
\langle S^2_{\rm eik} \rangle=\frac{\int {\rm d}^2 b_{1\perp}\,{\rm d}^2  b_{2 \perp}\, |\tilde{T}(s, b_{1\perp}, b_{2\perp})|^2\,{\rm exp}(-\Omega_{A_1 A_2}(s,b_\perp))}{\int {\rm d}^2\, b_{1\perp}{\rm d}^2 b
_{2\perp}\, |\tilde{T}(s,b_{1\perp},b_{2\perp})|^2}\;,
\end{equation}
where $b_{i\perp}$ is the impact parameter vector of ion $i$, so that $b_\perp=b_{1\perp}+ b_{2\perp}$ corresponds to the transverse separation between the colliding ions.  $\tilde{T}(s, b_{1\perp},b_{2\perp})$ is the amplitude \eqref{eq:tq1q2} in impact parameter space, i.e.
\be
\tilde{T}(s, b_{1\perp},b_{2\perp})=\frac{1}{(2\pi)^4} \int {\rm d}^2 q_{1\perp} {\rm d}^2 q_{2\perp} e^{-i \vec{q}_{1\perp}\cdot \vec{b}_{1\perp}}e^{i \vec{q}_{2\perp}\cdot \vec{b}_{2\perp}}T(s, q_{1\perp},q_{2\perp})\;,
\ee
while $\Omega_{A_1 A_2}(s,b_\perp)$ is the ion--ion opacity; physically $\exp(-\Omega_{A_1 A_2}(s,b_\perp))$ represents the probability that no inelastic scattering occurs at impact parameter $b_\perp$. Its calculation is described in the following section. For our purposes it is simpler to work in $q_\perp$ space, for which we introduce the screening amplitude via 
\be\label{eq:tres}
T_{\rm res}(q_{1\perp},q_{2\perp}) = \frac{i}{s} \int \frac{{\rm d}^2 k_\perp}{8\pi^2} T_{\rm el}(k^2_\perp) T(q_{1\perp}',q_{2\perp}')\;,
\ee
where $q_{1\perp}' =q_{\perp} - k_\perp$ and $q_{2\perp}' = q_{2\perp} + k_\perp$ and $T_{\rm el}$ is the elastic ion--ion amplitude, given by
\be
T_{\rm el}(k^2_\perp) = 2is \int {\rm d}^2 b_\perp\, e^{i \vec{k}_\perp \cdot \vec{b}_\perp}(1-e^{-\Omega_{A_1A_2}(b_\perp)/2})\;.
\ee
Then it is straightforward to show that 
\be
\langle S^2_{\rm eik} \rangle= \frac{{\rm d}^2 q_{1\perp}{\rm d}^2 q_{2\perp}|T(q_{1\perp},q_{2\perp}) + T_{\rm res}(q_{1\perp},q_{2\perp})|^2}{{\rm d}^2 q_{1\perp}{\rm d}^2 q_{2\perp}|T(q_{1\perp},q_{2\perp})|^2}\;,
\ee
and thus we should simply replace $T(q_{1\perp},q_{2\perp}) \to T(q_{1\perp},q_{2\perp}) + T_{\rm res}(q_{1\perp},q_{2\perp})$ for the corresponding amplitude in \eqref{eq:csn}.

\subsection{The ion--ion opacity}\label{sec:opac}

\begin{figure}
\begin{center}
\includegraphics[scale=0.35]{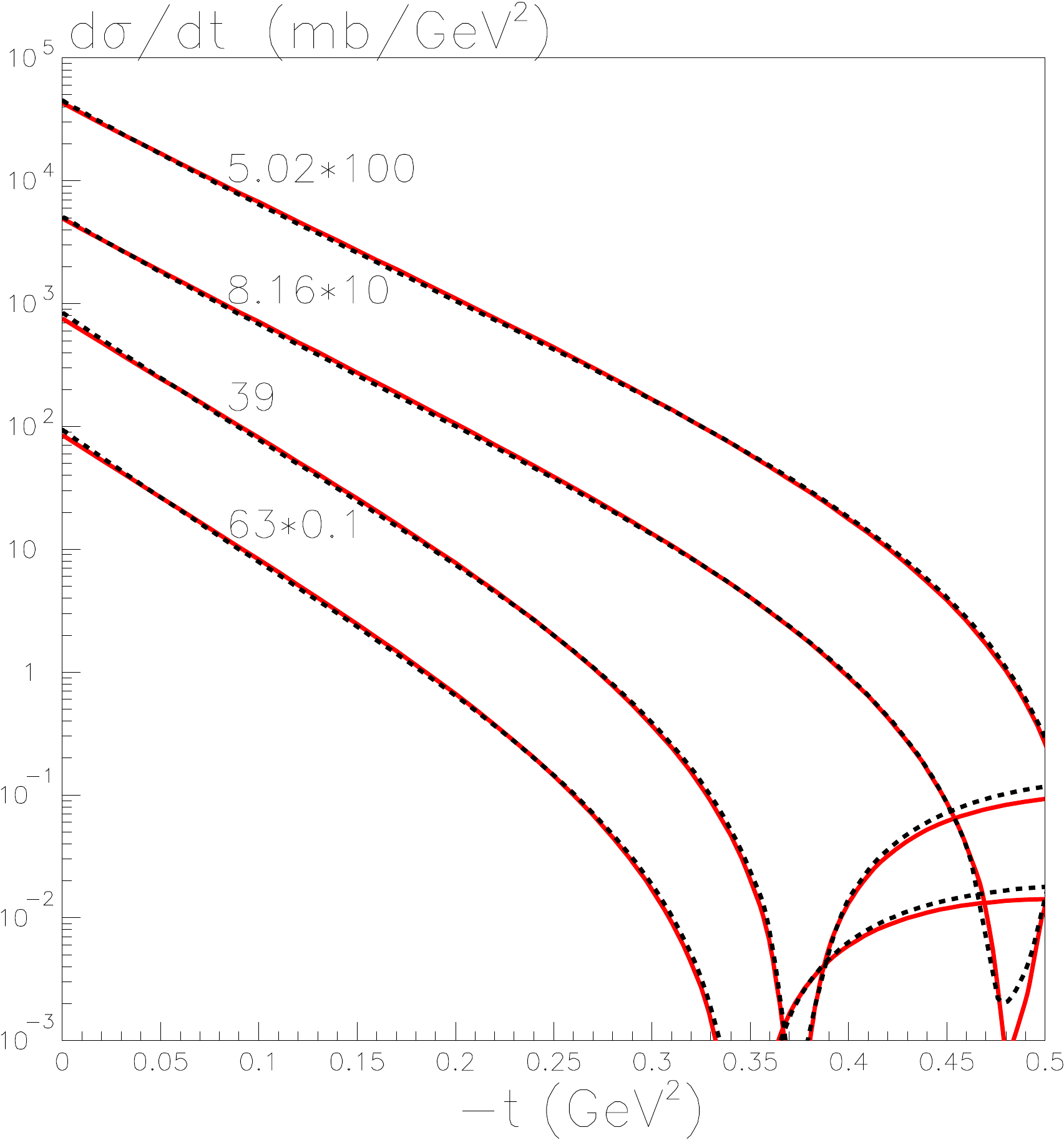}
\caption{Elastic proton--proton cross section $d\sigma/dt$ at 5,02, 8.16, 39 and 63 TeV (from top to bottom). The predictions calculated within the two--channel model~\cite{Khoze:2018kna} and the one channel eikonal model described in the text are shown by the red and dashed black lines, respectively. In both cases only the $|{\rm Im} A_{el}|^2$ contribution to $d\sigma/dt$ is shown.}
\label{fig:sigel}
\end{center}
\end{figure} 

Having introduced the ion--ion opacity above, which encodes the probability for no additional ion--ion rescattering at different impact parameters, we must describe how we calculate this.
The ion--ion opacity is given in terms of the opacity due to nucleon--nucleon interactions, $\Omega_{nn}$, which is in turn given by a convolution of the nucleon--nucleon scattering amplitude $A_{nn}$ and the transverse nucleon densities $T_n$. In particular we have 
\be\label{eq:omega}
\Omega_{A_1A_2}(b_\perp) = \int {\rm d}^2 b_{1\perp} {\rm d}^2 b_{2\perp} T_{A_1}(b_{1\perp})T_{A_2}(b_{2\perp})A_{nn}(b_\perp-b_{1\perp}+b_{2\perp})\;,
\ee
with $T_A$ given in terms of the nucleon density 
\be\label{eq:tpn}
T_A(b_\perp)= \int {\rm d}z\, \rho_A(r) = \int {\rm d}z\, (\rho_n(r) + \rho_p(r))\;,
\ee
of the corresponding ion. For the case of $pA$ collisions, we simply take
\be
T_A(b_\perp) \to \delta^{(2)}(\vec{b}_\perp)\;,
\ee
for the $A \to p$ replacement. The  nucleon--nucleon scattering amplitude is given in terms of the nucleon opacity $\Omega_{nn}(b_\perp)$  via
\be\label{eq:anncoh}
A_{nn}(b_\perp) = 2 ( 1-e^{-\Omega_{nn}(b_\perp)/2})\;.
\ee
Note that this corresponds to the total scattering cross section, as
\be
\sigma_{\rm tot}^{nn} = \int {\rm d}^2 b_\perp A_{nn}(b_\perp)\;,
\ee
see e.g.~\cite{Ryskin:2009tj}. This is the appropriate choice  the momentum transfers involved  even in purely elastic nucleon--nucleon rescattering will as a rule lead to ion break up. On the other hand for the case of QCD--initiated semi--exclusive production discussed further below, where the ion breaks up, we should take
\be\label{eq:annincoh}
A_{nn}(b_\perp) = 1-e^{-\Omega(b_\perp)}\;,
\ee
so that
\be
\sigma^{nn}_{\rm inel} = \int {\rm d}^2 b_\perp A_{nn}(b_\perp)\;,
\ee
which corresponds to a somewhat smaller suppression. To calculate the nucleon opacity we can then apply precisely the same procedure as for $pp$ collisions, see e.g.~\cite{Khoze:2013jsa,Khoze:2014aca}. This in general requires the introduction of so--called Good--Walker eigenstates~\cite{Good:1960ba} to account for the internal structure of the proton. However, in order to avoid unfeasibly complicated combinatorics we instead apply a simpler one--channel approach here. The parameters of this model are tuned in order to closely reproduce the more complete result of the two--channel model of~\cite{Khoze:2018kna} for the elastic $pp$ cross section in the relevant lower $t$ region, in particular before the first diffractive dip. The result is shown in Fig.~\ref{fig:sigel}.

In more detail, the nucleon opacity is given by
\be
\Omega(b_\perp) = -\frac{i}{s}\frac{1}{4\pi^2}\int  {\rm d}^2 q_\perp\, e^{i \vec{q}_\perp \cdot\vec{b}_\perp} A_{I\!\! P}(-q^2)\;,
\ee
where $A_{I\!\! P}$ is the elastic amplitude due to single Pomeron exchange, given by
\be
A_{I\!\! P}=is \sigma_0 \beta^2(t)\;.
\ee
For the form factors $\beta$ we take 
\be\label{eq:beta}
\beta(t)=\exp{(-(b(a-t))^c +  (ab)^c)}\;,
\ee
with the precise numerical values given in Table~\ref{tab:op} (for other values of $\sqrt{s}$ we use a simple interpolation). We note that in the above, we have the same scattering amplitude in the neutron and protons cases, due the high energy nature of the interaction and dominance of Pomeron exchange in this region.

\begin{table}
\begin{center}
\begin{tabular}{|c|c|c|c|c|}
\hline
$\sqrt s$ [TeV]& $\sigma_0$ [mb]& $a$  [GeV$^2$ ]& $b$ [GeV$^{-2}$] & c\\
\hline
 5.02 & 146 &       0.180 &        20.8   &    0.414 \\    
   8.16&    159    &   0.190     &   26.3   &    0.402\\    
   39 &    228 &       0.144   &     23.3     &  0.397\\    
   63 &   245    &   0.150     &   28.0    &   0.390\\  
   \hline  
\end{tabular}
\caption{The parameters of the one channel eikonal description of nucleon--nucleon amplitude, described in the text.} \label{tab:op}
\end{center}
\end{table}

\begin{figure}
\begin{center}
\includegraphics[scale=0.6]{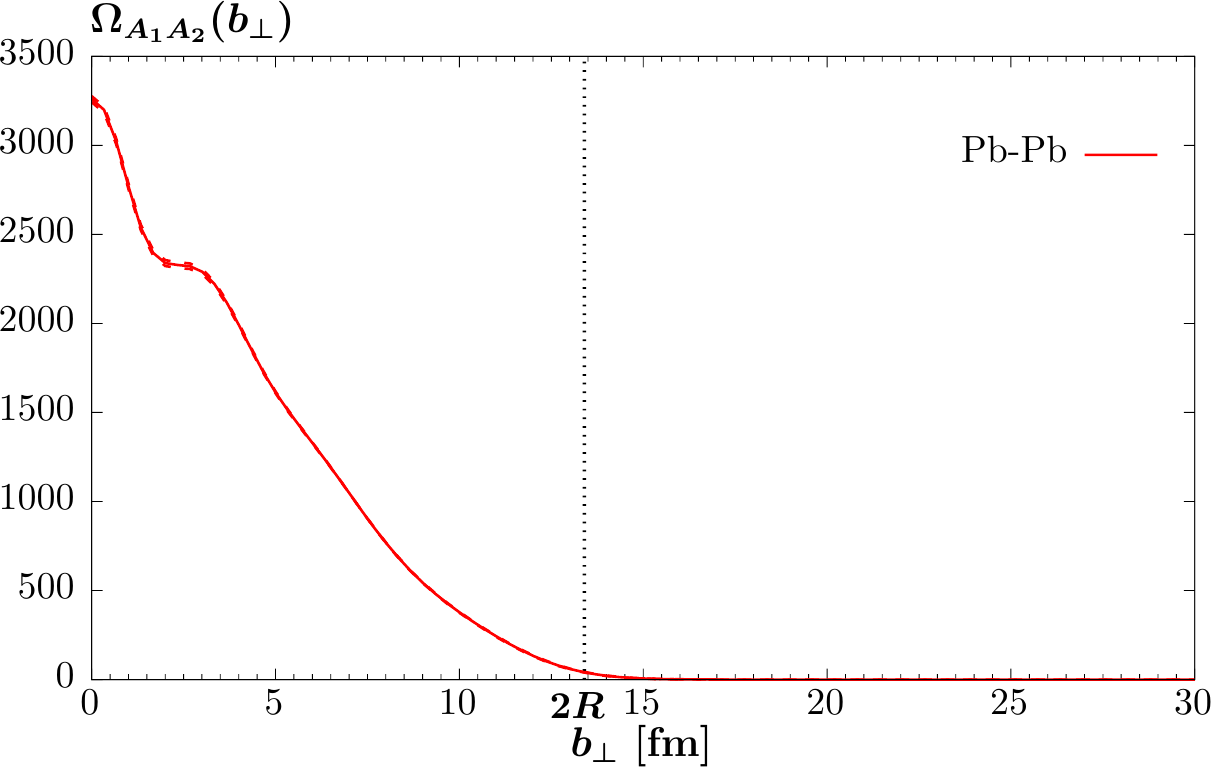}
\includegraphics[scale=0.6]{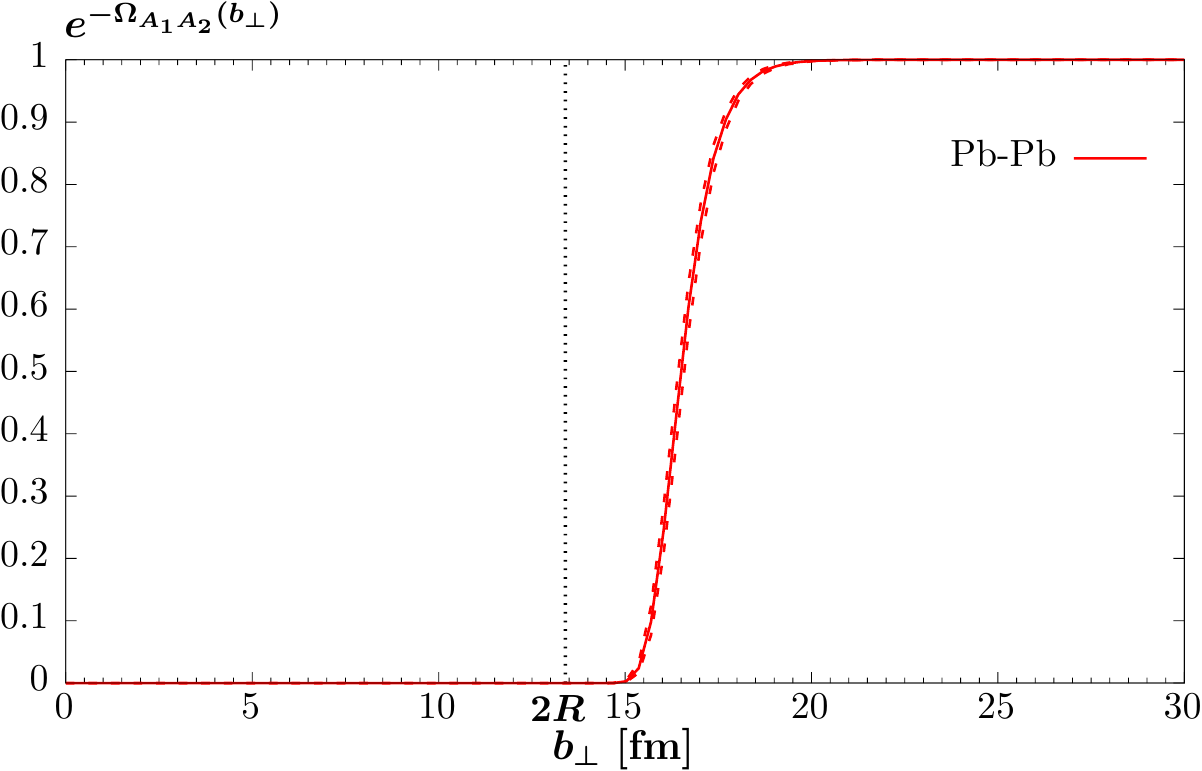}
\caption{Ion--ion opacity (left) and probability for no inelastic scattering (right) for lead--lead collisions, as a function of the lead impact parameter $b_\perp$.}
\label{fig:opacb}
\end{center}
\end{figure}

The opacity and probability for no inelastic scattering, $e^{-\Omega_{A_1 A_2}(b_\perp)}$, in lead--lead collisions are shown in Fig.~\ref{fig:opacb}. For the neutron and proton densities we take as before the Wood--Saxons distribution~\eqref{eq:rhop}, with the experimentally determined values~\cite{Tarbert:2013jze}
\begin{align}\nonumber
R_p &= 6.680\, {\rm fm}\;, &d_p &= 0.447 \, {\rm fm}\;,\\ \label{eq:pbpar}
R_n &= (6.67\pm 0.03)\, {\rm fm}\;, &d_n &= (0.55 \pm 0.01) \, {\rm fm}\;.
\end{align}
The solid curve corresponds to the central values, while for the dashed curves we take values for the neutron density at the lower and upper end of the 1$\sigma$ uncertainties, for illustration. For lower values of $b_\perp \lesssim 2 R$ (here we define $R\sim R_{p,n}$ for simplicity), where the colliding ions are overlapping in impact parameter space, we can see that the probability is close to zero, while for larger $b_\perp \gtrsim 2 R$ this approaches unity, as expected. However we can see that this transition is not discrete, with the probability being small somewhat beyond $2 R$, due both to the non--zero skin thickness of the ion densities and range of the QCD single--Pomeron exchange interaction. This will be missed by an approach that is often taken in the literature, namely  to simply to cutoff the cross in impact parameter space when $b_\perp<2R$. Comparing to \eqref{eq:S2},  we can see that this corresponds to taking instead
\be
e^{-\Omega(b)/2} = \theta(b-2 R)\;.
\ee
The value at which this would turn on is indicated in Fig.~\ref{fig:opacb}. As our more realistic result turns on smoothly above $2R$, this will correspond to somewhat suppressed exclusive cross sections in comparison. For ultra--peripheral photon-initiated interactions, where the dominant contribution to the cross section comes from $b_\perp \gg 2 R$, this will have a relatively mild impact, but for QCD--initiated production a complete treatment is essential.

\subsection{QCD--induced production}\label{sec:qcdind}

We can also apply the above formalism to the case of QCD--initiated diffractive production in heavy ions. We will discuss two categories for this, namely semi--exclusive and fully exclusive production, below.

\subsubsection{Semi--exclusive production}\label{sec:semiex}

We first consider the case of incoherent QCD--induced CEP. Here, while the individual nucleons remain intact due to the diffractive nature of the interaction, the ion will in general break up. This can therefore lead to an exclusive--like signal in the central detector, with large rapidity gaps between the produced state and ion decay products. If zero degree calorimeter (ZDC) detectors are not used to veto on events where additional forward neutrons are produced, this will contribute to the overall signal.

Nonetheless, as we will see such interactions are strongly suppressed by the requirement that the ions themselves do not interact in addition, producing secondary particles in the central detector, that is due the ion--ion survival factor. The incoherent cross section, prior to the inclusion of survival effects, is simply given by integrating the CEP cross section in $pp$  collisions over the nucleon densities 
\be\label{eq:sigincoh0}
\sigma_{\rm incoh} = \int {\rm d}^2 b_{1\perp} {\rm d}^2 b_{2\perp} T_{A_1}(b_{1\perp}) T_{A_2}(b_{2\perp}) \sigma_{\rm CEP}^{pp} \;,
\ee
where $\sigma_{\rm CEP}^{pp}$ is the usual QCD--induced $pp$ cross section as implemented in previous versions of \texttt{SuperChic}~\cite{Harland-Lang:2015cta}. Note that here we make the approximation that the nucleon--nucleon CEP interaction is effectively point--like in comparison to the ion radius $R$. Strictly speaking, we should instead convolute the transverse densities, $T_A$, with the form factor due to the range of the nucleon--nucleon CEP interaction, however we have checked that numerically this is a relatively small effect, and omit this in what follows. We can see that \eqref{eq:sigincoh0} then scales like $\sim A_1 A_2$, i.e. with the total number of nucleon pairings. However, this exclude survival effects. To account for these, we simply multiply by the probability for no additional inelastic ion--ion interactions, so that 
\be\label{eq:sigincoh}
\sigma_{\rm incoh} = \int {\rm d}^2 b_{1\perp} {\rm d}^2 b_{2\perp} T_{A_1}(b_{1\perp}) T_{A_2}(b_{2\perp}) \sigma^{\rm CEP} e^{-\Omega_{A_1 A_2}( b_{1\perp} - b_{2\perp})}\;,
\ee
where  the opacity $\Omega_{A_1 A_2}$  is calculated as described in Section~\ref{sec:opac}, in particular via~\eqref{eq:omega} and~\eqref{eq:annincoh}. In fact the MC, we calculate the effective survival factor
 \be
 \left \langle S^2_{\rm incoh} \right \rangle  = \frac{\int {\rm d}^2 b_{1\perp} {\rm d}^2 b_{2\perp} T_n(b_{1\perp}) T_n(b_{2\perp}) e^{-\Omega_{A_1 A_2}(b_{1\perp}-b_{2\perp})}}{\int {\rm d}^2 b_{1\perp} {\rm d}^2 b_{2\perp} T_n(b_{1\perp}) T_n(b_{2\perp}) }\;,
 \ee
 and multiply the usual $pp$ cross section (calculated differentially in $k_\perp$ space) by this.
 
Crucially, we have seen in Section~\ref{sec:opac}, see in particular Fig.~\ref{fig:opacb}, that the ion--ion opacity is very large for $b_\perp \lesssim 2 R$, and hence the probability of no additional inelastic interactions is exponentially suppressed. This has the result that in the region of significant nucleon density (where the $T_A$ are not suppressed) we will almost inevitably have additional inelastic interactions, and the corresponding survival factor will be very small. Thus, we will only be left with a non--negligible CEP cross section in the case that the interacting nucleons are situated close to the ion periphery, where the nucleon density and hence inelastic interaction probability is lower. In other words, we do not have $A_1 A_2$ possible nucleon--nucleon interactions, but rather expect a much gentler increase with the ion mass number, with only those nucleons on the surface (or more precisely, the edge of the ion `disc' in the transverse plane) playing a role. We find in particular that to good approximation
\be\label{eq:incohsc}
\sigma_{\rm incoh} \propto A^{1/3}\;,
\ee
for both $AA$ and $pA$ collisions, where in the former case we assume the ions are the same for simplicity.
The detailed derivation is given in Appendix~\ref{ap:incoh}. Comparing to the $Z_1^2 Z_2^2$ scaling of the photon--initiated process, we may therefore expect QCD--initiated CEP to be strongly suppressed; we will see that this is indeed the case below.

\subsubsection{Exclusive production}\label{sec:exprod}

Alternatively, we can consider the case of coherent ion--ion QCD--induced CEP, which leaves the  ions intact. To achieve this, we proceed in a similar way to Sections~\ref{sec:gamunsc} and~\ref{sec:gamsc}. In particular we simply have
\be\label{eq:tqcdex}
T_{\rm QCD}^{A_1 A_2}(q_{1\perp},q_{2\perp}) = T_{\rm QCD}^{pp}(q_{1\perp},q_{2\perp}) F_{A_1}(Q_1^2) F_{A_2}(Q_2^2)\;,
\ee
where $Q^2$ is given as in (\ref{eq:qi}) and $T_{\rm QCD}$ is the QCD--induced CEP amplitude as calculated within the usual Durham model approach, see~\cite{Harland-Lang:2015cta}. Here $F_A$ is the ion form factor, given in terms of the nucleon density $\rho_A$, see \eqref{eq:tpn}. In impact parameter space this corresponds to 
\be
\tilde{T}_{\rm QCD}^{A_1 A_2 }(b_{1\perp},b_{2\perp}) = \int {\rm d}^2 b_{1\perp}'{\rm d}^2 b_{2\perp}' \tilde{T}_{\rm QCD}^{pp}(b_{1\perp}',b_{2\perp}') T_{A_1}(b_{1\perp}-b_{1\perp}') T_{A_2}(b_{2\perp}-b_{2\perp}') \;.
\ee
Now, the range of the nucleon--nucleon CEP amplitude $T^{pp}$ (which is $\lesssim 1$ fm) is significantly less than the extent of the ion transverse density (i.e. $\sim 7$ fm for a Pb ion). This allows us to take $T_A(b_{\perp}-b_{\perp}') \sim T_A(b_{\perp})$ above, so that
\begin{align}
\tilde{T}_{\rm QCD}^{A_1 A_2 }(b_{1\perp},b_{2\perp}) &\approx T_{A_1}(b_{1\perp})T_{A_2}(b_{2\perp})\int {\rm d}^2 b_{1\perp}'{\rm d}^2 b_{2\perp}' \tilde{T}_{\rm QCD}^{pp}(b_{1\perp}',b_{2\perp}')\;,\\
& =  T_{\rm QCD}^{pp}(q_{1\perp}=0,q_{2\perp}=0)\cdot T_{A_1}(b_{1\perp})T_{A_2}(b_{2\perp})\;.
\end{align}
The cross section then becomes
\begin{align}\nonumber
\sigma_{\rm coh} &= (4\pi^2)^2 |T_{\rm QCD}^{pp}(q_{1\perp}=0,q_{2\perp}=0)|^2 \int {\rm d}^2 b_{1\perp}{\rm d}^2 b_{1\perp}|T_{A_1}(b_{1\perp})|^2 |T_{A_2}(b_{2\perp})|^2 e^{-\Omega_{A_1 A_2}( b_{1\perp} - b_{2\perp})}\;,\\ \label{eq:sigbt}
&\approx  (4\pi^2)^2 \frac{\sigma_{\rm CEP}^{pp}}{\pi^2 \left\langle q_{1\perp}^2 \right \rangle \left\langle q_{2\perp}^2 \right \rangle }\int {\rm d}^2 b_{1\perp}{\rm d}^2 b_{1\perp}|T_{A_1}(b_{1\perp})|^2 |T_{A_2}(b_{2\perp})|^2 e^{-\Omega_{A_1 A_2}( b_{1\perp} - b_{2\perp})}\;,
\end{align}
where we define $\left\langle q_{\perp}^2 \right \rangle$ in the second line. This is of the order of the average squared transverse momentum transfer in the $pp$ cross section, i.e.
\be
\left\langle q_{1\perp}^2 \right \rangle \sim \frac{\int {\rm d} q_{1\perp}^2 q_{1\perp}^2 |T_{\rm QCD}^{pp}(q_{1\perp},q_{2\perp})|^2}{\int{\rm d} q_{1\perp}^2 |T_{\rm QCD}^{pp}(q_{1\perp},q_{2\perp})|^2}\;,
\ee
and similarly for $q_{2\perp}$, where we assume the $q_{1\perp}$ and $q_{2\perp}$ dependencies factorise; such an expression is exactly true if we assume a purely exponential form factor in $q_\perp^2$, for example.

We emphasise that in the MC we make use of the general result, with the formalism of Section~\ref{sec:gamsc} applied to \eqref{eq:tqcdex} to include survival effects. However, the above result \eqref{eq:sigbt} holds to good approximation, and allows us to derive some straightforward expectations for the scaling and size of the coherent contribution. As discussed further in Appendix~\ref{ap:incoh}, under these approximations, for $pA$ collisions we expect a similar $\sim A^{1/3}$ to the incoherent case~\eqref{eq:incohsc}, but with a parametric suppression 
\be\label{eq:cohsupp}
\sigma_{\rm coh}^{pA} \sim \frac{4\pi}{\sigma_{\rm tot}^{nn} \left \langle q_{\perp}^2 \right \rangle} \sigma_{\rm incoh}^{pA}  \;.
\ee
For $AA$ collisions the expected scaling is in fact somewhat gentler in comparison to the incoherent case, with a (squared) parametric suppression
\be\label{eq:cohsup}
\sigma_{\rm coh}^{AA} \sim \left(\frac{4\pi}{\sigma_{\rm tot}^{nn} \left \langle q_{\perp}^2 \right \rangle}\right)^2A^{-1/6}\cdot \sigma_{\rm incoh}^{AA}  \propto A^{1/6}\;,
\ee
where we write $\left \langle q_{i\perp}^2 \right \rangle = \left\langle q_\perp^2 \right \rangle$ and $\sigma_{\rm tot}^{nn}$ is the total $pp$ cross section.
We therefore expect some numerical parametric suppression by the ratio of the cross sectional extent of the CEP interaction with each ion ($\sim 4\pi/\left\langle q_\perp^2 \right \rangle$) to total $pp$ cross section. Taking some representative vales for these, numerically we have 
\be\label{eq:qtsub}
\frac{4\pi}{\sigma_{\rm tot}^{nn} \left \langle q_{\perp}^2 \right \rangle} \sim \frac{4\pi}{90 \,{\rm mb}  \cdot 0.1 \,{\rm GeV}^2} \sim 0.5\;.
\ee
Hence we may expect some suppression in the coherent cross section, although given the relatively mild effect predicted by this approximate result, a precise calculation is clearly necesssary. Note that we here take a rather small value of $\left\langle q_{\perp}^2 \right \rangle \sim 0.1$ ${\rm GeV}^2$, corresponding to a quite steep slope in $q_\perp^2$. This is as expected when $pp$ rescattering effects are included, see e.g.~\cite{HarlandLang:2010ep}, which tend to prefer small values of the proton transverse momenta, where the survival factor is larger. A consequence of this is that the observed ratio of the cross section with heavy ions to the proton--proton cross section will depend on the precise process considered and in particular the quantum numbers of the produced state, through the effect this has on the survival factor.

Finally, we recall that in the case of ion--ion collisions there is a reasonable probability to excite a `giant dipole resonance' (GDR) via multi--photon exchange between the ions. This effect, not currently included in the MC, will lead to an excited final state, decaying via the emission of additional neutrons. From~\cite{Baltz:2002pp}, the probability for this to occur at the relative low impact parameters $b_\perp \sim 2R$ relevant to QCD--initiated CEP is found to be rather large, see Fig.~2 of this reference. We can estimate from this a probability of $\sim$ 50\% for GDR excitation in each ion in this region. This will reduce the exclusive and increase the semi--exclusive cross sections predicted here accordingly. If one does not tag neutron emission experimentally via ZDCs this is not an issue, as we simply sum the two contributions, however when comparing to data with such tagging performed a corresponding correction to our predictions should be made.

\subsubsection{Including the participating nucleons}\label{sec:part}

In principle our calculation of the survival factor in proton--ion and ion--ion collisions, as in for example \eqref{eq:sigincoh}, i.e.
\be
S^2_{\rm A_1 A_2} (b_\perp)  = e^{-\Omega_{A_1 A_2}( b_{\perp} )}\;,
\ee
gives the probability of no inelastic interactions between all nucleons within the overlap in impact parameter of the colliding ions. In particular, this corresponds to a simple Poissonian no interaction probability, with the mean number of inelastic nucleon--nucleon interactions given as in \eqref{eq:omega}, in terms of the total ion transverse densities $T_A$ integrated over the appropriate impact parameter regions. These therefore in principle take care of all possible nucleon--nucleon interactions, including the particular nucleon--nucleon pairing that undergoes CEP. 

However, the survival factor due to this active pair would be better treated separately and included explicitly, as its precise value will depend on the underlying CEP process. More significantly, the exclusive production process must take place close to the peripherary of the ions, where the corresponding nucleon density is low and the average number of nucleon--nucleon interactions contained in the above expression can be below one. Applying the above factor alone will therefore overestimate the corresponding survival factor, giving a value higher than that due to the active pair, and so such a separate treatment is essential.

We therefore include the (process dependent) nucleon--nucleon survival factor explicitly, i.e. the CEP cross section in \eqref{eq:sigincoh} and amplitude in \eqref{eq:tqcdex} correspond to those including survival effects in the nucleon--nucleon interaction. On the other hand, having done this we must take care to avoid double counting the possibility for inelastic interactions due to this active pair. Unfortunately this in general requires a careful treatment of the ion structure, moving beyond the opacity above, which is simply given in terms of the total average nucleon density. Here, we base our calculation on the nuclear shell model, and recall that for CEP we are dominated by interactions which occur close to the ion peripherary, which is mainly populated by $N_{\rm shell}$ nucleons with the largest principal and orbital quantum numbers. Each of these contributes to the total average nucleon density
\be
T_A(b_\perp) = \sum_{i=1}^{N_{\rm shell}} T_A^i(b_\perp) = N_{\rm shell}  T_A^i(b_\perp) \;,
\ee
where $T_A^i(b_\perp)$ is the contribution from each individual nucleon, which in the last step we assume to be the same for each nucleon. To remove the contribution from the active nucleon that undergoes CEP we therefore simply replace
\be
\Omega_{pA} \to \Omega_{pA} \left(1-\frac{1}{N_{\rm shell}}\right)\;,\qquad \Omega_{AA} \to \Omega_{AA} \left(1-\frac{1}{N_{\rm shell}}\right)^2\;,
\ee
in the corresponding opacities. In the case of $^{208}Pb$ the highest shell has $l=3$ for neutrons and $l=2$ for protons, corresponding to 14 neutrons and 12 protons. At the peripherary the proton density is roughly three times smaller than the neutron, and therefore as a rough estimate then we can take $N_{\rm shell} \approx 20$. Hence this correction is rather small, at the $5-10\%$ level.

However, this is not the end of the story. In particular the position of the nucleons in the ion shell are not completely independent, and  we can expect some repulsion between them due to $\omega$ meson exchange~\cite{Reid:1968sq}. In the ion peripherary the nucleon density is rather small, and hence it is reasonable to describe this repulsion in the same way as the repulsive `core' in the deuteron wave function~\cite{Reid:1968sq}. Here, the separation between the nearest nucleons cannot be less than  $r_{\rm core}=0.6-0.8$ fm. To account for this, we can subtract an interval of length $2r_{\rm core}$ in the $z$ direction from the nucleon density \eqref{eq:tpn} which enters the calculation of the opacity\footnote{To be precise, we omit the region $(-r_{\rm core},r_{\rm core})$, that is we take mean value of $z=0$ for the active nucleon. In the case of the ion--ion opacity, for which additional nucleon--nucleon interactions can take place at different impact parameters to the active nucleon, such a simple replacement will in general overestimate the cross section, but for peripheral collisions this remains a good approximation.}.
 
 In the results which follow we will take $N_{\rm shell}=20$ and $r_{\rm core}=0.8$ fm. The latter gives roughly a $50\%$ increase in the cross section, while as discussed above the former correction is significantly smaller. While this provides our best estimate of the CEP cross section, there is clearly some uncertainty in the precise predictions due to the effects above, conservatively at the $50\%$ level, with the result omitting these two corrections representing a lower bound on the cross section.

\subsubsection{Numerical results}

\begin{figure}
\begin{center}
\includegraphics[scale=0.6]{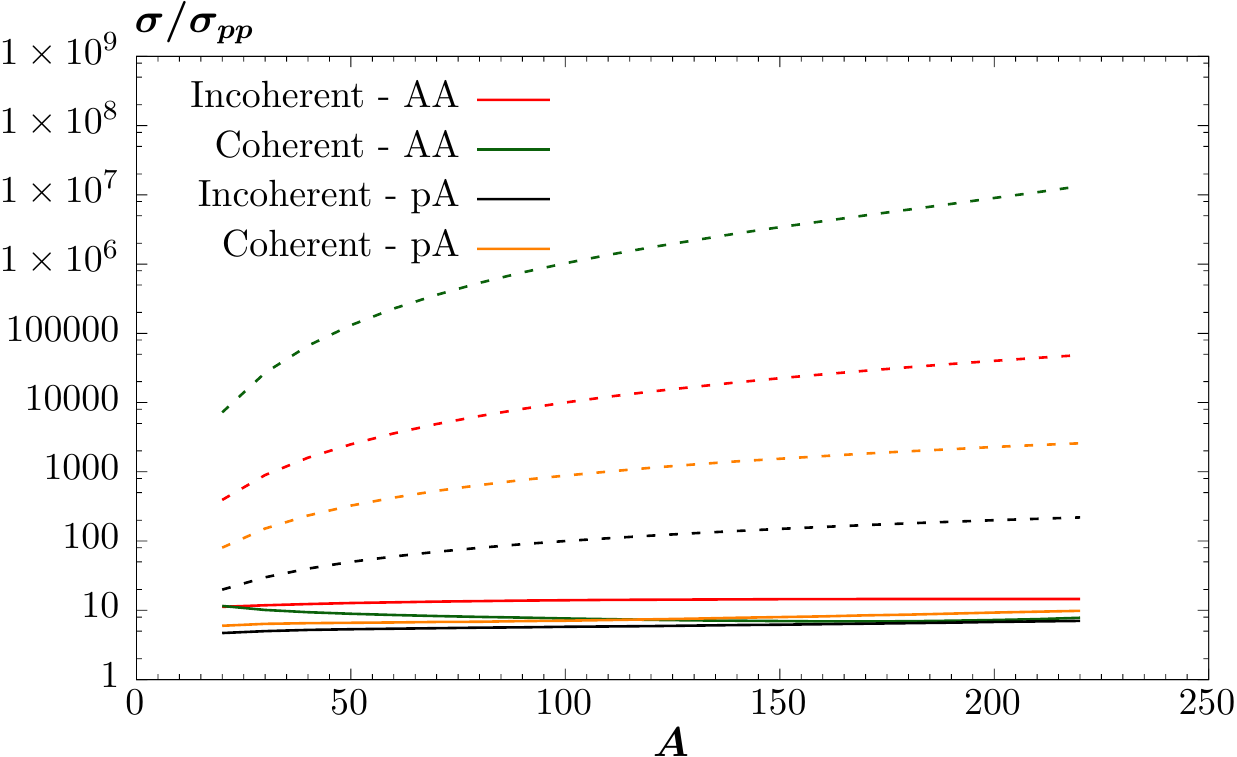}
\includegraphics[scale=0.6]{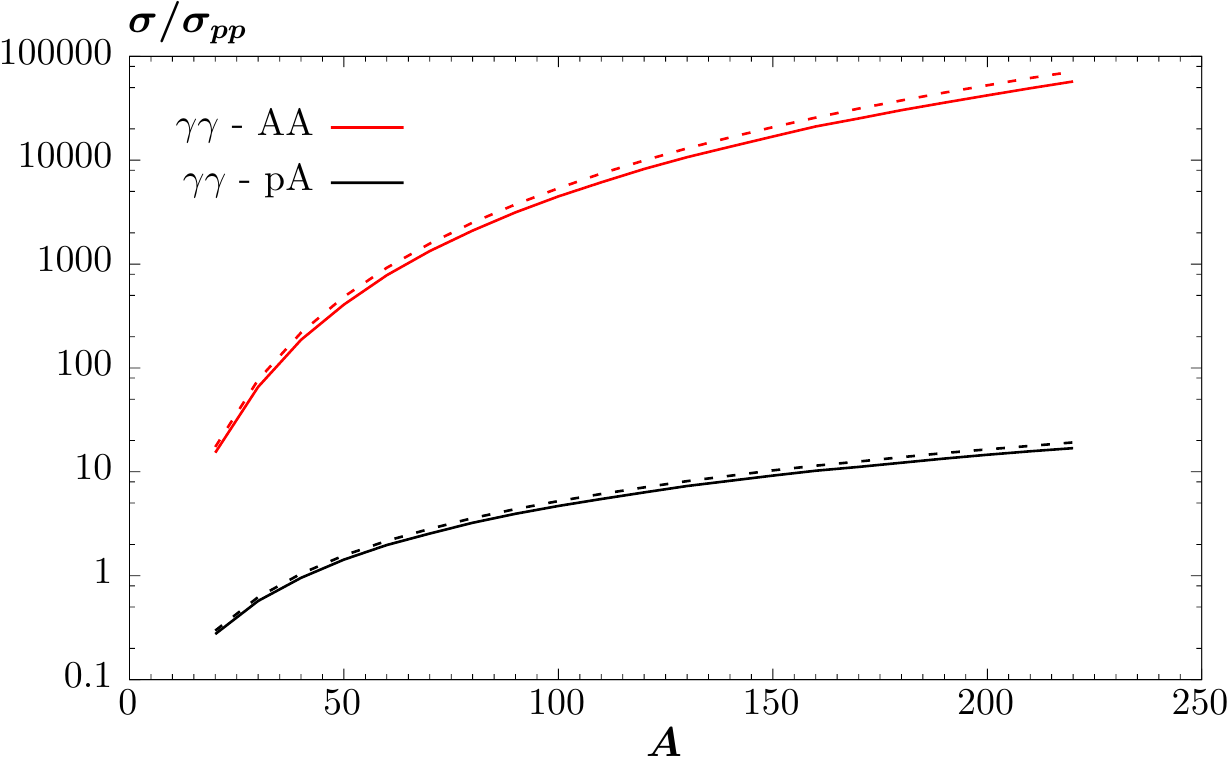}
\caption{Ratio of cross sections at $\sqrt{s}=5.02$ TeV in proton--ion ($pA$) and ion--ion ($AA$) collisions to the proton--proton result. The QCD (photon) initiated cases are shown in the left (right) plots. Results with and without survival effects are shown by the solid (dashed) lines. Note that for the QCD--initiated production the survival factor due to the participating nucleon pair is included in all cases.}
\label{fig:qcdcomp}
\end{center}
\end{figure}

\begin{figure}
\begin{center}
\includegraphics[scale=0.66]{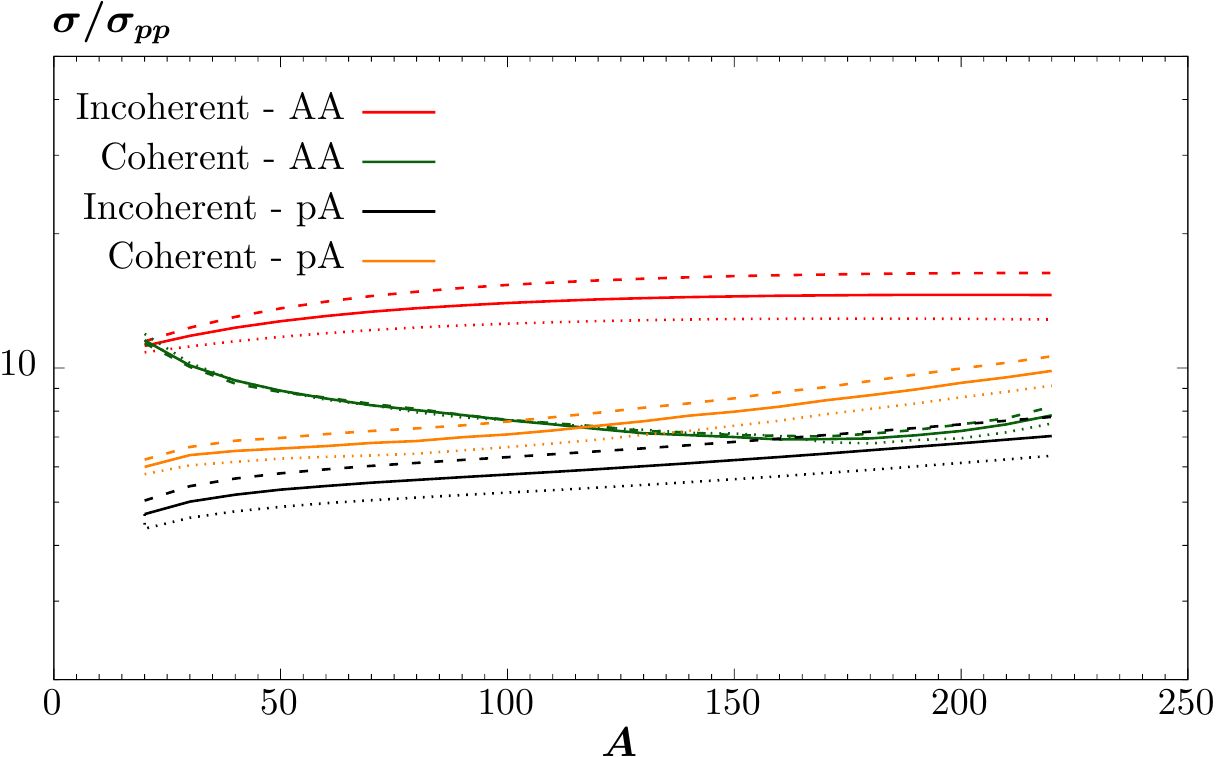}
\caption{Ratio of QCD--initiated cross sections at $\sqrt{s}=5.02$ TeV in proton--ion ($pA$) and ion--ion ($AA$) collisions to the proton--proton result. Results are shown with different values of the ion skin thickness, as described in the text.}
\label{fig:qcdcomp1}
\end{center}
\end{figure}

In Fig.~\ref{fig:qcdcomp} (left) we show numerical predictions for the ratio of QCD--initiated cross sections at $\sqrt{s}=5.02$ TeV in proton--ion ($pA$) and ion--ion ($AA$) collisions to the proton--proton result. In all cases we include the survival factor due to the active nucleon pair, but in the solid curves we include the effect due to the additional nucleons present in the ion(s) as well. To be concrete, we show results for $\gamma\gamma$ production within the ATLAS event selection~\cite{Aaboud:2017bwk}.  We take \eqref{eq:rdep} for the dependence of the ion radius on $A$, while we show results for $d=0.5, 0.55$ and 0.6 fm (dotted, solid and dashed lines, respectively), including survival effects, in Fig.~\ref{fig:qcdcomp1} to give an indication of the sensitivity of the cross section to the value of the ion skin thickness. This also provides a clearer demonstration of the trends for the full cross section (i.e. including survival effects): the solids curves in the two plots correspond to the same results. 

In all cases, the impact of survival effects is found to be sizeable. Already for proton--ion collisions these reduce the corresponding cross sections by up to two order of magnitude, while in ion--ion collisions the effect is larger still, leading to a reduction of up to four and six orders of magnitude in the semi--exclusive and exclusive cases, respectively. As discussed earlier, this is to be expected: as the range of the QCD--initiated CEP interaction is much smaller than the ion radius, the majority of potential nucleon--nucleon CEP interactions (in the absence of survival effects due to the non--interacting nucleons) would take place in a region of high nucleon density, where additional particle production is essentially inevitable. This is in strong contrast to the case of photon--initiated production, where the long range QED interaction allows all protons in the ion to contribute coherently in an ultra--peripheral process.

Considering in more detail the cross sections including survival effects, in the proton--ion case, the relatively gentle scaling of the exclusive and semi--exclusive cross sections with $A$ is clear, which upon inspection are indeed found to follow a rough $\sim A^{1/3}$ trend, consistent with \eqref{eq:incohsc} and \eqref{eq:cohsupp}. 
As expected from the discussion in Section~\ref{sec:exprod}, the exclusive and semi--exclusive cross sections are of similar sizes. Interestingly, we can see that the precise calculation predicts that the exclusive cross section is in fact somewhat enhanced relative to the semi--exclusive.
For the ion--ion case we can see that the semi--exclusive cross section again increases only very gently with $A$, again as expected. Upon inspection, we observe that the trend is consistent with a flatter $A$ dependence then the simple $\sim A^{1/3}$ scaling predicted using the analytic calculation of Appendix~\ref{ap:incoh}; on closer investigation, we find that this is due to the correct inclusion of the impact parameter dependence of the elastic nucleon--nucleon scattering amplitude in the definition of the opacity~\eqref{eq:omega}, which is omitted in the simplified analytic approach. In the exclusive case, interestingly the cross section in fact decreases with $A$, albeit with a relatively flat behaviour at larger $A$. This is again found to be due to the full calculation of the opacity. 
Again, numerically the exclusive and semi--exclusive cross sections enter at roughly the same order, with some suppression in the former case, as expected  from the discussion in Section~\ref{sec:exprod}.

In Fig.~\ref{fig:qcdcomp} (right) we show the corresponding cross section ratios for the photon--initiated cross sections. For concreteness, we calculate
$Z$ by maximising the binding energy according to the semi--empirical mass formula~\cite{Weizscker:1935zz}, i.e.
\be
\frac{A}{Z}\approx 2+\frac{a_C}{2 a_A}A^{2/3}\;,
\ee
with $a_C=0.711$, $a_A=23.7$. The impact of survival effects is in this case found to be significantly more moderate, at the $10 -20\%$ level, due to the well--known result that the photon--initiated interaction takes places at large impact parameters, i.e. ultra--peripherally, where the impact of further ion--ion or proton--ion interactions is relatively small. The dramatic cross section scaling with $A$ in the ion--ion case is also clear, leading to a relative enhancement by many orders of magnitude in comparison to the QCD--initiated case. For proton--ion collisions a milder enhancement is also observed. We note that in both cases the steeply falling $Q^2$ dependence of the ion form factors leads to some suppression relative the na\"{i}ve $\sim Z^2$ and $Z^4$ scaling in the proton--ion and ion--ion cases.

\section{New processes}\label{sec:newproc}

In this section we briefly describe the new processes and refinements that have been included in \texttt{SuperChic} since the version described in~\cite{Harland-Lang:2015cta}.

\subsection{Light--by--light scattering: $W$ loop contributions}

In previous versions of \texttt{SuperChic}, expressions for the fermion loop contributions to the  $\gamma \gamma \to \gamma \gamma$ light--by--light scattering process in the $\hat{s} \gg m_f^2$ limit were applied. We now move beyond this approximation, applying the \texttt{SANC} implementation~\cite{Bardin:2009gq} of this process, which includes the full dependence on the fermion mass in the loop. This in addition includes the contribution from $W$ bosons, which was not included previously, again with the full mass dependence. We also implement a modified version of the  \texttt{SANC} implementation for the $gg \to \gamma\gamma$ process, which has the same form as the quark--loop contributions to the light--by--light scattering process, after accounting for the different colour factors and charge weighting. 

\begin{figure}
\begin{center}
\includegraphics[scale=0.3]{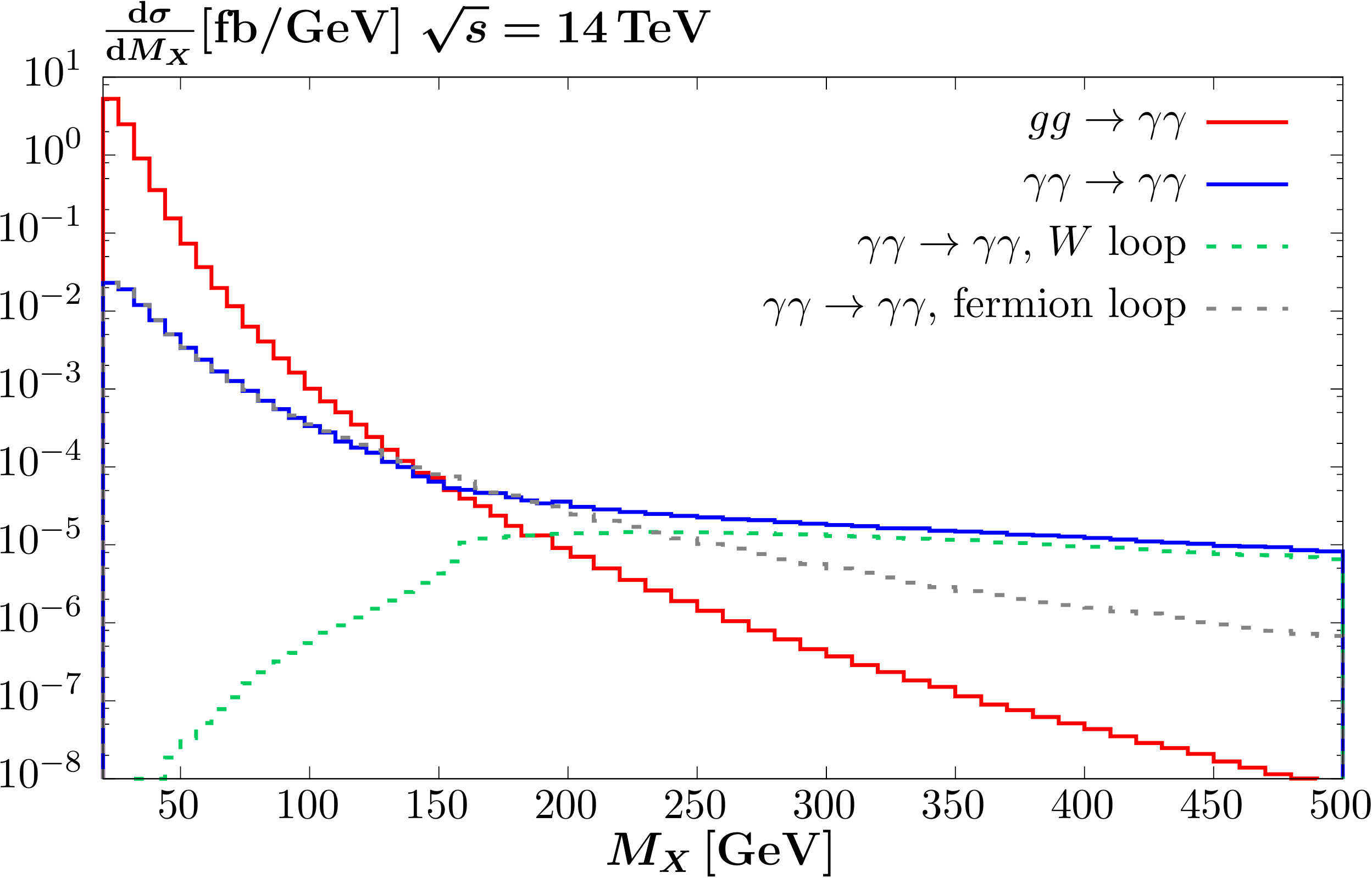}
\includegraphics[scale=0.64]{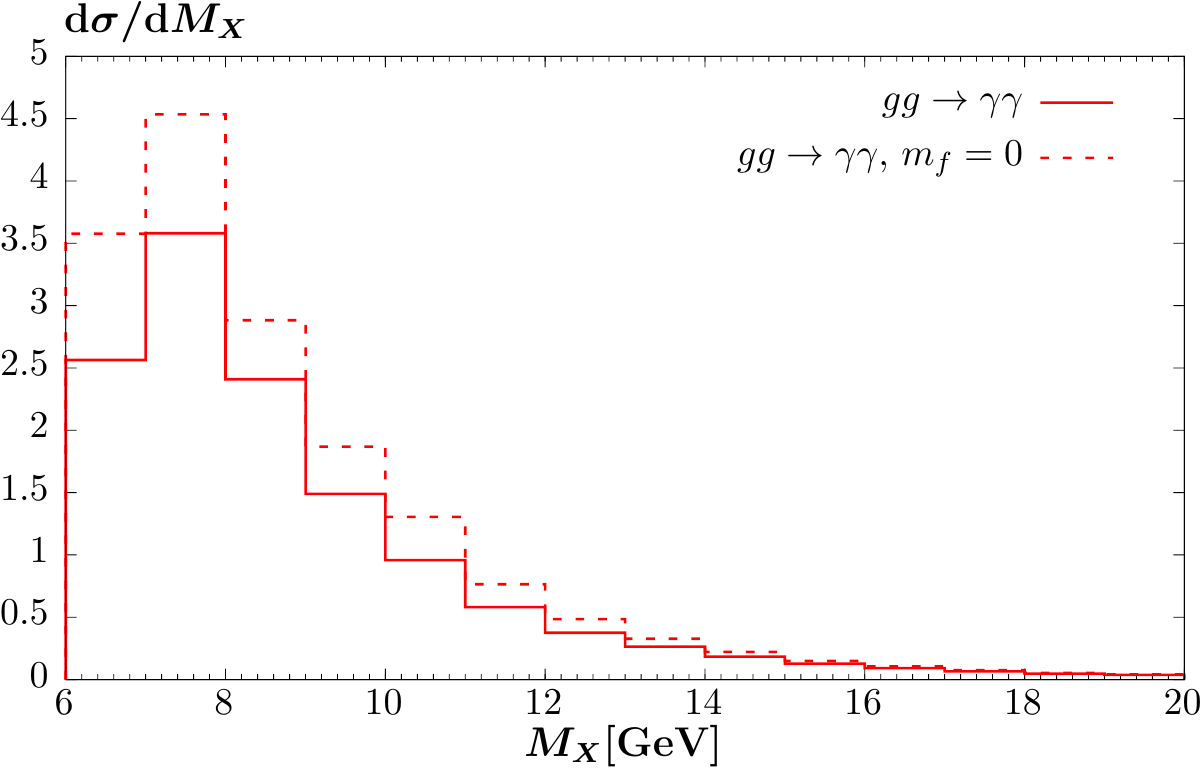}
\caption{Diphoton invariant mass distribution due to QCD and photon--initiated CEP in $pp$ collisions at $\sqrt{s}=14$ TeV. The left plot in addition shows the individual contributions from fermion and $W$ loops to the $\gamma\gamma$--initiated process, while the right plot shows the impact of including finite fermion masses.}
\label{fig:lbyl}
\end{center}
\end{figure}

In Fig.~\ref{fig:lbyl} (left) we show the diphoton invariant mass distribution due to QCD and photon--initiated CEP in $pp$ collisions at $\sqrt{s}=14$ TeV. The photons are required to have transverse momentum $p_\perp^\gamma > 10$ GeV and pseudorapidity $|\eta^\gamma|<2.4$. We can see that while the former dominates for $M_{\gamma\gamma} \lesssim 150$ GeV, above this the latter is more significant. This is due to the well--known impact of the Sudakov factor in the QCD--initiated cross section~\cite{Khoze:2001xm} which suppresses higher mass production, due to the increasing phase space for additional gluon radiation, so that at high enough mass this compensates the suppression in the photon--initiated cross section due to the additional powers of the QED coupling $\alpha$. We also show the relative contributions of fermion and $W$ boson loops to the photon--initiated cross section. While for $M_{\gamma\gamma} \lesssim 2 M_W$ the latter is as expected negligible, at sufficiently high invariant mass it comes to dominate. In Fig.~\ref{fig:lbyl} (right) we show the impact of excluding the fermion masses for the QCD--initiated case. The photons are required to have transverse momentum $p_\perp^\gamma > 16$ GeV and pseudorapidity $|\eta^\gamma|<2.4$. We can see that at lower $M_X$ the difference is at the $\sim 30\%$ level, decreasing to below $10\%$ at higher mass, in the considered region. Thus the previous \texttt{SuperChic} predictions will have overestimated the cross section by this amount. It should be noted however, that for the $gg\to \gamma\gamma$ case this is below the level of other theoretical uncertainties, due in particular to the gluon PDF and soft survival factor. Moreover, this is a purely LO result, and we may expect higher order corrections to increase the cross section by a correction of this order.

Finally, we note that the MC prediction for QCD--initiated CEP processes such as diphoton production does not include the impact of so--called `enhanced'  screening effects. These may be expected to reduce the corresponding cross section by as much as a factor of $\sim 2$~\cite{Ryskin:2009tk,Ostapchenko:2017prv}, but we leave a detailed study of this to future work. Note that such effects are entirely absent in the case of photon--initiated CEP.

\subsection{ALP production}

New light pseudoscalar `axion--like' particles (ALPs), with dimension--5 couplings to two gauge bosons or derivative interactions to fermions occur in a wide range of BSM models, often resulting from the breaking of some approximate symmetry (for a list of popular references, see e.g.~\cite{Baldenegro:2018hng}). For example, in the context of dark matter, these are often considered as mediators between dark matter and SM particles, while from an observational point of view the coupling to the SM may be sufficiently small so as to evade current constraints. The production of ALPs in ultra--peripheral heavy ion collisions was discussed in~\cite{Knapen:2016moh}, and more recently in~\cite{Baldenegro:2018hng} for the case of larger ALP masses, in $pp$ collisions, while the ATLAS evidence for light--by--light scattering~\cite{Aaboud:2017bwk} was used in~\cite{Knapen:2017ebd} and in the recent CMS analysis~\cite{dEnterria:2018uly} to set the most stringent constraints yet on the ALP mass and couplings in certain regions of parameter space.

\begin{figure}
\begin{center}
\includegraphics[scale=0.66]{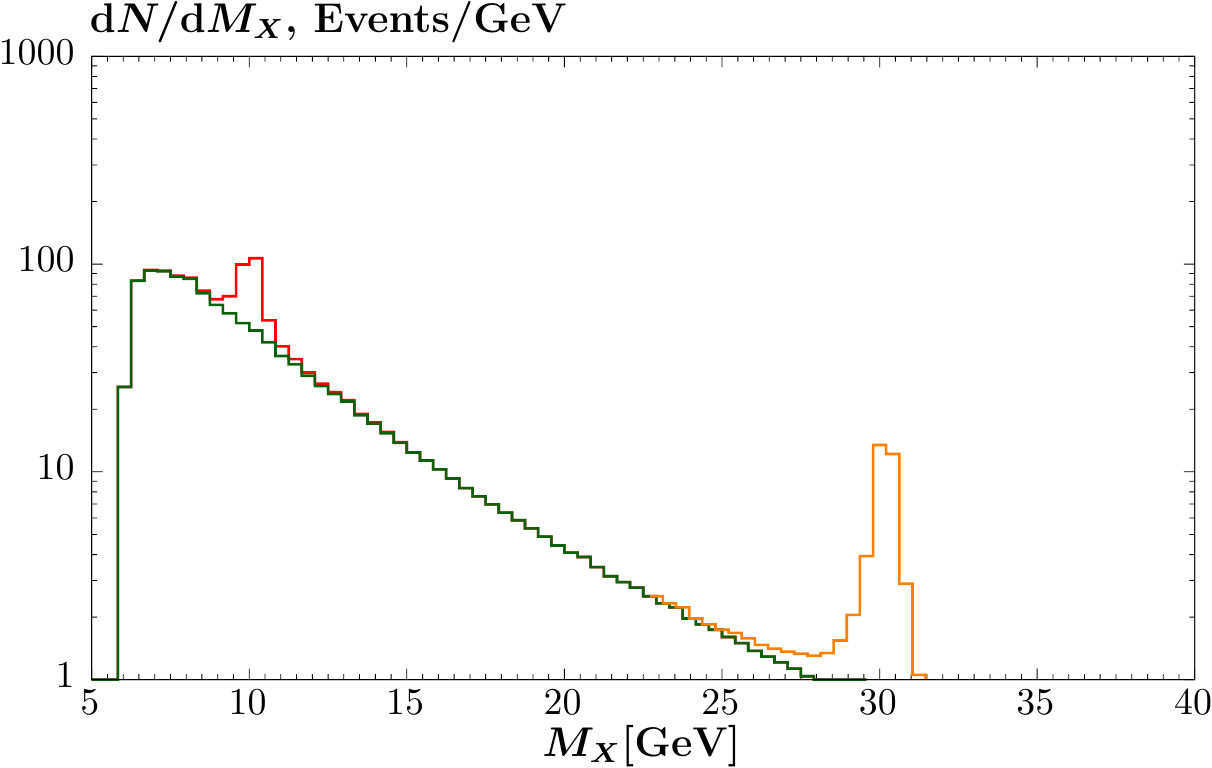}
\caption{Diphoton invariant mass distribution at $\sqrt{s}=5.02$ TeV in Pb--Pb collisions, for integrated luminosity $L=10\,{\rm nb}^{-1}$. The result due to the production of an ALP of mass 10 and 30 GeV is shown, with coupling $g_a=5\times 10^{-5}$ ${\rm GeV}^{-1}$, in both cases with a width of $0.5$ GeV included to roughly mimic the effect of experimental resolution. The continuum light--by--light background is also shown. The photons are required to have transverse momentum $p_\perp^\gamma>3$ GeV and pseudorapidity $|\eta^\gamma|<2.4$. The ALP is assumed here only to couple to photons.}
\label{fig:alp}
\end{center}
\end{figure}

We implement ALP production according to the Lagrangian
\be
\mathcal{L}=\frac{1}{2}\partial^\mu a \partial_\mu a -\frac{1}{2}m_a^2 a^2 -\frac{1}{4}g_a a F^{\mu\nu}\tilde{F}_{\mu\nu}\;,
\ee
where $\tilde{F}^{\mu\nu}  = \frac{1}{2}\epsilon^{\mu\nu \alpha\beta} F_{\alpha\beta}$. That is, we only consider $\gamma\gamma$ coupling with strength $g_a$, through which the ALP is both produced and decays. We in addition include the possibility of a scalar ALP, through the replacement $\tilde{F} \to F$. For the $\gamma_{\lambda_1}\gamma_{\lambda_1} \to a$ amplitudes these give:
\begin{align}
{\rm Pseudoscalar}&: & & \mathcal{M}_{+-}=\mathcal{M}_{-+}=0\;,& & \mathcal{M}_{++}=-\mathcal{M}_{--}=\frac{g_a M_{\gamma\gamma}^2}{2}\;,\\
{\rm Scalar}&: & & \mathcal{M}_{+-}=\mathcal{M}_{-+}=0\;,& & \mathcal{M}_{++}=\mathcal{M}_{--}=\frac{g_a M_{\gamma\gamma}^2}{2}\;.
\end{align}
As an example, the expected signals due to a 10 and 30 GeV pseudoscalar ALP, with coupling $g_a=5\times 10^{-5}$ ${\rm GeV}^{-1}$, are shown in Fig.~\ref{fig:alp}, overlaid on the continuum light--by--light background. The expected number of events (ignoring any further experimental efficiencies) with $L=10\,{\rm nb}^{-1}$ of  $\sqrt{s}=5.02$ TeV Pb--Pb collision data are shown. We note that in both cases these are not excluded by current experimental constraints~\cite{Knapen:2017ebd,dEnterria:2018uly}.

\subsection{Monopole and monopolium production}

Magnetic monopoles complete the symmetry of Maxwell's equations and explain charge quantization~\cite{Dirac:1931kp}. As such states would be expected to have large electromagnetic couplings, one possibility is to search for the production of monopole pairs, or bound states of monopole pairs (so called `monopolium') through exclusive photon--initiated production at the LHC~\cite{Epele:2012jn}. In the MC we have implemented the CEP of both monopoles pairs, and monopolium, in the latter case followed by the decay to two photons.

For the production of monopole pairs, we simply apply the known results for lepton pair production $\gamma\gamma \to l^+l^-$, but with the replacement $\alpha \to 1/4\alpha$, as required by the Dirac quantisation condition
\be
g=N\frac{2\pi}{e}\;,
\ee
where we take $N=1$, and $g$ is the monopole charge. We also allow for the so--called  $\beta g$  coupling scenario~\cite{Epele:2012jn}, for which we simply replace $g \to g \beta$, 
 where $\beta$ is the monopole velocity. In the monopolium case we apply the cross section of~\cite{Epele:2012jn}, with the wave function of~\cite{Epele:2007ic}, and include the decay to two photons.

\subsection{$\gamma\gamma \to t\overline{t}$}

We include photon--initiated top quark production. This is implemented using the same matrix elements as the lepton pair production process, with the mass, electric charge and colour factors suitably modified. We find a total photon--initiated cross section of $0.25$ fb in $pp$ collisions at $\sqrt{s}=14$ TeV, and 36 fb in Pb--Pb collisions at $\sqrt{s}=5.02$ TeV. Note that the QCD--initiated cross section in $pp$ collisions is about $0.02$ fb, and so is an order of magnitude smaller, while in Pb--Pb this will be smaller still.

\section{Light--by--light scattering: a closer look}\label{sec:LbyL}

Evidence for light--by--light scattering in ultra--peripheral Pb--Pb collisions has been found by ATLAS~\cite{Aaboud:2017bwk} and more recently by CMS~\cite{dEnterria:2018uly}. In both cases, the production of a diphoton system accompanied by no additional particle production is measured, while in the ATLAS case ZDCs are in addition used to measure additional neutral particle production in the forward direction, which would be a signal of semi--exclusive production accompanied by ion break--up. 

\begin{table}
\begin{center}
\begin{tabular}{|c|c|c|c|c|}
\hline
&LbyL&QCD (coh.) & QCD (incoh.) & $A^2 R^4$\\
\hline
ATLAS & 50 &0.008 &0.05 &50\\
\hline
ATLAS (aco $<$ 0.01, $p_\perp^{\gamma\gamma}<2$ GeV) & 50& 0.007&0.01 &10\\
\hline
CMS &103 &0.03 &0.2&180\\
\hline
CMS (aco $<$ 0.01, $p_\perp^{\gamma\gamma}<1$ GeV) &102 & 0.02&0.03&30\\
\hline
\end{tabular}
\caption{Predicted cross sections, in nb, for diphoton final states within the ATLAS~\cite{Aaboud:2017bwk} and CMS~\cite{dEnterria:2018uly} event selections, in Pb--Pb collisions at $\sqrt{s}=5.02$ TeV. That is, the photons are required to have transverse energy $E_\perp^\gamma >2$ (3) GeV and pseudorapidity $|\eta^\gamma|<2.4$, while in the CMS case an additional cut of $m_{\gamma\gamma} > 5$ GeV is imposed. Results with and without an additional acoplanarity cut aco $<$ 0.01, and cut on the combined transverse momentum $p_\perp^{\gamma\gamma}<1$ (2) GeV in the CMS (ATLAS) case are shown. The cross sections for the light--by--light scattering (LbyL) and QCD--initiated photon pair production, in both the coherent and incoherent cases, are given. The result of simply scaling the $pp$ cross section (including the $pp$ survival factor) by $A^2 R^4$ with $R=0.7$ is also shown.} \label{tab:lbylcs}
\end{center}
\end{table}

However, in addition to the desired photon--initiated signal, there is the possibility that QCD--initiated diphoton production may contribute as a background. We are now in a position for the first time to calculate this, using the results of Section~\ref{sec:qcdind}. The results for the QCD--initiated background (both coherent and incoherent), as well as the prediction for the light--by--light signal, are shown in Table~\ref{tab:lbylcs}. We consider both the ATLAS and CMS event selection in the central detectors. Namely, the produced photons are required to have transverse energy $E_\perp^\gamma >2$ (3) GeV and pseudorapidity $|\eta^\gamma|<2.4$ in the case of CMS (ATLAS), while for CMS an addition cut of $m_{\gamma\gamma} > 5$ GeV is imposed.  We show results before and after further cuts on the diphoton system $p_\perp^{\gamma\gamma}<1 (2) $ GeV for CMS (ATLAS) and acoplanarity ($1-\Delta \phi_{\gamma\gamma}/\pi<0.01$) are imposed, which are designed to suppress the non--exclusive background.

For the light--by--light signal the predicted  cross sections are fully consistent with the ATLAS and CMS results:
\begin{align}
\sigma^{\rm ATLAS} &=70 \pm 24\, ({\rm stat.}) \pm 17 \,({\rm syst.}) \,{\rm nb} \;,\\
\sigma^{\rm CMS} &=120 \pm 46 \,({\rm stat.}) \pm 28\, ({\rm syst.})\pm 4\, ({\rm th.}) \,{\rm nb} \;.
\end{align}
On the other hand, we find that the QCD--initiated background is expected to be very small. In particular, both the incoherent and coherent contributions are expected to be negligible, even before imposing additional acoplanarity cuts.

\begin{figure}
\begin{center}
\includegraphics[scale=0.66]{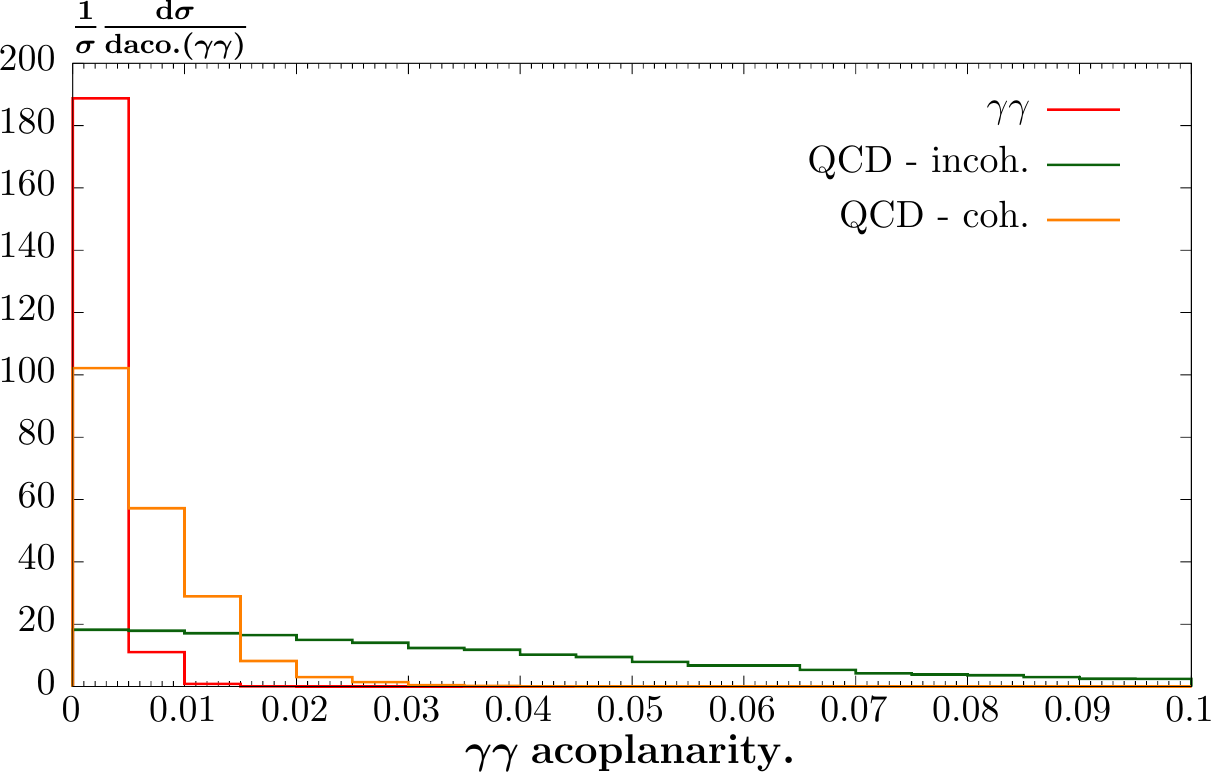}
\caption{Normalized differential cross sections for exclusive and semi--exclusive diphoton production with respect to the diphoton acoplanarity. The QED--initiated and QCD--initiated (both coherent and incoherent) processes are shown.}
\label{fig:aco}
\end{center}
\end{figure}

We can see that incoherent background, which we recall corresponds to the case that the colliding ions do not remain intact, is 
 further suppressed by the additional acoplanarity and $p_\perp^{\gamma\gamma}$ cuts; as we would expect, due to the broader $p_\perp$ spectrum of the incoherent cross section. This is seen more clearly in Fig.~\ref{fig:aco}, which shows the (normalized) acoplanarity distributions in the three cases. We can see that the QED--initiated process is strongly peaked at low acoplanarity ($<$ 0.01), as is the coherent QCD--initiated process, albeit with a somewhat broader distribution due to the broader QCD form factor in this case. On the other hand, for incoherent QCD--initiated production we can see that the spectrum is spread quite evenly over the considered acoplanarity region. 

It was suggested in~\cite{d'Enterria:2013yra} that to calculate the QCD--initiated background, understood to be the dominant incoherent part, we can simply scale the corresponding $pp$ cross section by a factor of $A^2 R^4$, where $R\approx 0.7$ accounts for nuclear shadowing effects. As discussed in Section~\ref{sec:qcdind}, this $\sim A^2$ scaling is certainly far too extreme, due to the short--range nature of the QCD interaction and corresponding requirement that only peripheral interactions can lead to exclusive or semi--exclusive production. In addition, we note that as the dominant contribution in this case will come from nucleons situated close to the ion peripherary, where the nucleon number density is relatively low, we can expect shadowing effects to be minimal, and hence we are justified in using the standard proton PDF in the calculation of the CEP cross section. Nonetheless, for the sake of comparison we also show the predictions from this $\sim A^2 R^4$ scaling in Table~\ref{tab:lbylcs}, where we include the $pp$ survival factor. We can see that the cross section prediction in this case is, as expected, much larger, by many orders of magnitude. Such an approach will therefore dramatically overestimate the expected background.  On the other hand, the relative reduction with the application of the acoplanarity and $p_\perp^{\gamma\gamma}$  cuts is similar to the semi--exclusive case, being driven by the same QCD form factor which enters in both cases.

It should be emphasised that in both the ATLAS and CMS analyses the normalization of the QCD--initiated background is in fact determined by the data. In particular, the predicted QCD background from this $A^2 R^4$ scaling is allowed to be shifted by a free parameter $f^{\rm norm}$, which is fit to the observed cross section in the aco $>0.01$ region, where the LbyL signal is very low.  Interestingly in both analyses a value of $f^{\rm norm} \approx 1$ is preferred, which is significantly larger than our prediction; from Table~\ref{tab:lbylcs} we can roughly expect $f^{\rm norm}  \sim \sigma^{\rm incoh}/\sigma^{A^2 R^4} \sim 10^{-3}$. However, great care is needed in interpreting these results: as is discussed in~\cite{dEnterria:2018uly} this normalization effectively account for {\it all} backgrounds that result in large acoplanarity photons, not just those due to QCD--initiated production. Indeed, in this analysis it is explicitly demonstrated that the MC for the background for $e^+ e^-$ production significantly undershoots the data in the large acoplanarity region, and it is suggested that this could be due to events where extra soft photons are radiated. Our results clearly predict that the contribution to the observed events in the large acoplanarity region should not be due to QCD--initiated production, suggesting that a closer investigation of other backgrounds, such as the case of $e^+ e^- + \gamma$ discussed in~\cite{dEnterria:2018uly}, would be worthwhile.

Finally, we note that in the ATLAS analysis~\cite{Aaboud:2017bwk} the number of events in the region with diphoton acoplanarity $> 0.01$, where the QED--initiated CEP signal will be strongly suppressed, with and without neutrons detected in the ZDCs is observed. They find 4 events with a ZDC signal, that is with ion dissociation, and 4 without, which roughly corresponds to a $O(10\,{\rm fb})$ cross section in both cases. However from Table~\ref{tab:lbylcs} we predict a much smaller cross sections of roughly $0.04$ (0.01) fb with (without) ZDC signals, i.e. 0 events in both cases. While some care is needed, in particular as the predictions in Table~\ref{tab:lbylcs} have not been corrected for detector effects, this predicted QCD contribution is clearly far too low to explain these observed events. We note that the probability of excitation of a GDR in each ion can be rather large (in~\cite{Baltz:2002pp} a probability of $\sim 30\%$ for the related vector meson photoproduction process is predicted), however these should generally lead to events in the acoplanarity $<0.01$ region. Inelastic photon emission can lead to ion break up at larger acoplanarity, but is predicted in~\cite{Hencken:1995me} to be at the \% level. Again, clearly further investigation of these issues is required.

\section{\texttt{SuperChic 3}: generated processes and availability}\label{sec:superchic}

\texttt{SuperChic 3} is a Fortran based Monte Carlo that can generate the processes described above and in~\cite{Harland-Lang:2015cta}, with and without soft survival effects. User--defined distributions may be output, as well as unweighted events in the HEPEVT, Les Houches and HEPMC formats. The code and a user manual can be found at http://projects.hepforge.org/superchic.

	Here we briefly summarise the processes that are currently generated, referring the reader to the user manual for further details. 
	The QCD--initiated production processes are: SM Higgs boson via the $b\overline{b}$ decay, $\gamma\gamma$, 2 and 3--jets, light meson pairs ($\pi,K,\rho,\eta('),\phi$), quarkonium pairs ($J/\psi$ and $\psi(2S)$) and single quarkonium ($\chi_{c,b}$ and $\eta_{c,b}$). Photoproduction processes are: $\rho$, $\phi$, $J/\psi$, $\psi(2S)$ and $\Upsilon(1S)$. Photon--initiated processes are: $W$ pairs, lepton pairs, $\gamma\gamma$, SM Higgs boson via the $b\overline{b}$ decay, ALPs, monopole pairs and monopolium. $pp$, $pA$ and $AA$ collisions are available for arbitrary ion beams, for QCD and photon--initated processes. For photoproduction, currently only $pp$ and $pA$ beams are included. Electron beams are also included for photon--initiated production.

\section{Conclusions and outlook}\label{sec:conc}

In this paper we have presented the updated \texttt{SuperChic 3} Monte Carlo generator for central exclusive production. In such a CEP process, an object $X$ is produced, separated by two large rapidity gaps from intact outgoing protons, with no additional hadronic activity. This simple signal is associated with a broad and varied phenomenology, from low energy QCD to high energy BSM physics, and is the basis of an extensive experimental programme that is planned and ongoing at the LHC.

\texttt{SuperChic 3} generates a wide range of final--states, via QCD and photon--initiated production and with pp, pA and AA beams. The addition of heavy ion beams is a completely new update, and we have included a complete description of both photon and QCD--initiated production. In the latter case this is to the best of our knowledge the first time such a calculation has been attempted. We have accounted for the probability that the ions do not interact inelastically, and spoil the exclusivity of the final state. While this is known to be a relatively small effect in the photon--initiated case, in the  less peripheral QCD--initiated case the impact has been found to be dramatic. 

These issues are particularly topical in light of the recent ATLAS and CMS observations of exclusive light--by--light scattering in heavy ion collisions. We have presented a detailed comparison to these results, and have shown that the signal cross section can be well produced by our SM predictions, and any background from QCD--initiated production is expected to be essentially negligible, in contrast to some estimates presented elsewhere in the literature. We find that the presence of additional events outside the signal region, with and without neutrons observed in the ZDCs (indicating ion break up) cannot be explained by the predicted QCD--initiated background. Addressing this open question therefore remains an experimental and/or theoretical challenge for the future.

Finally, there are very promising possibilities to use the CEP channel at high system masses to probe electroweakly coupled BSM states with tagged protons during nominal LHC running, accessing regions of parameters space that are difficult or impossible to reach using standard inclusive search channels. With this in mind, we have presented updates for photon--initiated production in pp collisions, including axion--like particle, monopole pairs and monopolium, as well as an updated calculated of SM light--by--light scattering including $W$ boson loops. These represent only a small selection of possible additions to the MC, and indeed as the programme of CEP measurements at the LHC continues to progress, we can expect further updates to come.

\section*{Acknowledgements}

We thank David d'Enterria and Marek Tasevsky for useful discussions, Vadim Isakov for useful clarifications on questions related to nuclear structure, and Radek {\v Z}leb{\v c}{\' a}k for identifying various bugs and mistakes in the previous MC version. LHL thanks the Science and Technology Facilities Council (STFC) for support via grant awards ST/P004547/1. MGR thanks the IPPP at the University of Durham for hospitality. VAK acknowledges  support from a Royal Society of Edinburgh  Auber award.

\appendix

\section{$A$ scaling in  QCD--induced production}\label{ap:incoh}

In this appendix we derive the scaling behaviour \eqref{eq:incohsc} for QCD--initiated production in heavy ion collisions. As discussed in Section~\ref{sec:semiex}, we are interested in the peripheral region, $r \gtrsim R$. We denote the direction of the ion--ion impact parameter $b_\perp$ as $x$ and the orthogonal transverse direction as $y$. We can write the $x$ position for each ion as $x_i   = R + \delta x_i$, with $i=1,2$. As we have $R\gg d$ we can expand in $\delta x/R$, to give
\be
r_i-R_i \approx \frac{y^2+z^2}{2R_i}+\delta x_i\;, 
\ee
where we have used that $y_1=y_2=y$ and $z_1=z_2=z$. We then have
\be\label{eq:apptb}
T(b_{i\perp}) = \int {\rm d}z\, \rho(r) \approx \rho_0 \int {\rm d}z\, e^{-\frac{r_i-R_i}{d}}\approx \rho_0 \int {\rm d}z \,e^{-\frac{y^2+z^2}{2R_i d}}e^{-\frac{\delta x_i}{d}}=\rho_0 \sqrt{2\pi R_i d}\,e^{-\frac{y^2}{2R_i d}}e^{-\frac{\delta x_i}{d}}\;.
\ee
In what follows, we will consider for simplicity a point--like QCD interaction. In other words, in the case of exclusive production for the ion--ion opacity we have 
\begin{align}
\Omega_{A_1A_2}(b_\perp) &= \int {\rm d}^2 b_{1\perp} {\rm d}^2 b_{2\perp} T_{A_1}(b_{1\perp})T_{A_2}(b_{2\perp})A_{nn}(b_\perp-b_{1\perp}+b_{2\perp})\;,\\
&\approx \sigma_{\rm tot}^{nn} \int {\rm d}^2 b_{1\perp}  T_{A_1}(b_{1\perp})T_{A_2}(b_{1\perp}-b_\perp)\;,
\end{align}
which is valid when the $nn$ interaction radius is much smaller than the extent of the ion transverse densities. In setting the normalization we have used \eqref{eq:anncoh}. For the case of semi--exclusive production, we simply replace $\sigma_{\rm tot}^{nn} \to \sigma_{\rm inel}^{nn}$, see~\eqref{eq:annincoh}. As we are only interested in the overall scaling with $A$, we will for simplicity assume $\sigma_{\rm inel}^{nn} \sim \sigma_{\rm tot}^{nn}$, and work with the latter variable in what follows; however, strictly speaking this replacement should be made when considering semi--exclusive production. We now consider the proton--ion and ion--ion cases in turn.

\subsection{Proton--ion collisions}

In this case, we take $T_{A_2} (b_\perp) = \delta^{(2)}(\vec{b}_{\perp})$, so that the opacity simply becomes
\be
\Omega_{pA}(b_\perp) = \sigma_{\rm tot}^{nn} T_A(b_\perp) \approx \sigma_{\rm tot}^{nn}\rho_0 (2\pi Rd)^{1/2} e^{-\frac{\delta x}{d}}\equiv \omega e^{-\frac{\delta x}{d}}\;,
\ee
which defines the constant $\omega$. Here, we have used the fact that for proton--ion collisions, the coordinate choice we have taken above corresponds to setting $y=0$, and we drop the subscript on the $\delta x$ for simplicity. Note that the integration is explicitly only performed over the peripheral region, i.e. over a ring of radius $\sim R$ and thickness $\delta x$, where we will expect a non--negligible contribution to the CEP cross section. Recalling \eqref{eq:sigincoh}, the incoherent cross section is given by 
\be
\sigma_{\rm incoh}^{pA} = \sigma_{\rm CEP}^{pp} \int {\rm d}^2 b_{\perp} T_A(b_{\perp}) e^{-\Omega(b_\perp)} = 2\pi R \frac{\omega}{\sigma_{\rm tot}^{nn}} \sigma_{\rm CEP}^{pp} \int {\rm d} \delta x \exp\left[ -\frac{\delta x}{d} -\omega e^{-\frac{\delta x}{d}}\right]\;.
\ee
The exponent falls sharply with increasing $\delta x$, and has a maximum at $\delta x = d \ln \omega$. We can therefore apply the saddle point approximation to evaluate the integral, giving
\be
\sigma_{\rm incoh}^{pA} \approx  2\pi R \frac{\omega}{\sigma_{\rm tot}^{nn}} \sigma_{\rm CEP}^{pp}\cdot  \frac{(2\pi)^{1/2}}{e} d \omega^{-1}  = \frac{(2\pi)^{3/2}}{e} \frac{Rd}{\sigma_{\rm tot}^{nn}} \sigma_{\rm CEP}^{pp}\;.
\ee
Taking $R \approx (4\pi/3)^{1/3} A^{1/3} \rho_0^{-1/3}$, we then have
\be
\sigma_{\rm incoh}^{pA} \approx \left(\frac{4\pi}{3}\right)^{1/3} \frac{(2\pi)^{3/2}}{e}\frac{d}{\rho_0^{1/3} \sigma_{\rm tot}^{nn}} \cdot A^{1/3} \cdot \sigma_{\rm CEP}^{pp}\sim 1.0 \cdot A^{1/3} \cdot \sigma_{\rm CEP}^{pp}\;,
\ee
where for concreteness we have substituted the values $\sigma_{\rm tot}^{nn}=90$ mb,  $\rho_0=0.15 \,{\rm fm}^{-3}$ and $d=0.5$ fm.

For coherent production, we have instead
\begin{align}
\sigma_{\rm coh}^{pA} &= \frac{4\pi}{ \left\langle q_{\perp}^2\right \rangle} \sigma_{\rm CEP}^{pp} \int {\rm d}^2 b_{\perp} T_A(b_{\perp})^2 e^{-\Omega(b_\perp)}\;,\\
&= \frac{4\pi}{ \left\langle q_{\perp}^2\right \rangle} \sigma_{\rm CEP}^{pp}  \frac{\omega^2}{(\sigma_{\rm tot}^{nn})^2}2\pi R \int {\rm d} \delta x \exp\left[ -2\frac{\delta x}{d} -\omega e^{-\frac{\delta x}{d}}\right]\;.
\end{align}
The exponent now has a maximum at $\delta x = d \ln \left(\omega/2\right)$, and we find 
\begin{align}
\sigma_{\rm coh}^{pA} &= \frac{4\pi}{ \left\langle q_{\perp}^2\right \rangle} 2\pi R \frac{\omega^2}{(\sigma_{\rm tot}^{nn})^2} \sigma_{\rm CEP}^{pp}\cdot \frac{4\pi^{1/2} d}{e^2} \omega^{-2} =\frac{4\pi}{ \left\langle q_{\perp}^2\right \rangle \sigma_{\rm tot}^{nn}} \frac{2^{3/2}}{e} \sigma_{\rm incoh}^{pA}\;,\\
& \sim \frac{4\pi}{ \left\langle q_{\perp}^2\right \rangle \sigma_{\rm tot}^{nn}} \sigma_{\rm coh}\sim 0.2 \cdot A^{1/3} \cdot \sigma_{\rm CEP}^{pp} \;,
\end{align}
where we have substituted numerically as in \eqref{eq:qtsub}.

\subsection{Ion--ion collisions}

For simplicity we will assume that $R_1=R_2=R$ in what follows, although the results can be readily be generalised. In this case, the opacity takes the form 
\be
\Omega_{AA}(b_\perp) =  \sigma_{\rm tot}^{nn} \int {\rm d}^2 b_{1\perp} T_A(b_{1\perp}) T_A(b_\perp-b_{1\perp})= \sigma_{{\rm tot}}\cdot  2\pi R d\rho_0^2 \int {\rm d}x {\rm d} y\, e^{-\frac{y^2}{R d}}e^{-\frac{\Delta}{d}}\;,
\ee
where we have imposed the constraint that $\delta x_1 + \delta x_2  = |b_\perp| - 2R \equiv \Delta$, which defines $\Delta$.
 Performing the integrals we have 
\be\label{eq:pcalc}
\Omega_{AA}(b_\perp) =\sigma_{\rm tot}^{nn}\cdot  2 (\pi R d)^{3/2}\rho_0^2\, \Delta e^{-\frac{\Delta}{d}} \equiv D \Delta e^{-\frac{\Delta}{d}}\;,
\ee
where we integrate $x$ over the interval $\Delta$. Considering first the incoherent cross section, we have
\be
\sigma_{\rm incoh}^{AA} = \frac{\sigma_{\rm CEP}^{pp}}{\sigma_{\rm tot}^{nn}}  \int {\rm d}^2 b_\perp \,P e^{-P}\;
\ee
where $P=D \Delta e^{-\Delta/d}$.

As before we only integrate over the peripheral region, with a ring of thickness $\Delta$ and radius $2R$. We have
\begin{align}\label{eq:incohaa}
\sigma_{\rm incoh}^{AA} &= 4\pi R \frac{\sigma_{\rm CEP}^{pp}}{\sigma_{\rm tot}^{nn}} \int {\rm d}\Delta \,P e^{-P}\;,\\
& = 4\pi R \frac{\sigma_{\rm CEP}^{pp}}{\sigma_{\rm tot}^{nn}}\int {\rm d}P\, \frac{e^{-P}}{|{\rm d}\ln P/{\rm d}\Delta|}\;,\\
& = 4\pi R \frac{\sigma_{\rm CEP}^{pp}}{\sigma_{\rm tot}^{nn}}\int {\rm d}P\, \frac{e^{-P}}{\left|\frac{1}{\Delta} - \frac{1}{d}\right|}\;.
\end{align}
The dominant contribution to this last integral comes from the region of $P\sim 1$. As an example, for the case of colliding lead ions, with $R=6.68$ fm, d=0.5 fm and $\sigma_{\rm tot}^{nn}=100$ mb for $\sqrt{s}=5.02$, we find $D\sim 15$ ${\rm fm}^{-1}$ in \eqref{eq:pcalc}. Thus $P \sim 1$ implies a rather large value of $\Delta \sim 1.5$ fm, i.e. $\Delta \sim 3 d$. This gives ${\rm d}\ln P/{\rm d}\Delta \sim -2/3d$ and hence the exclusive contribution comes from a ring in $b$ space of radius $R_1+R_2$ and thickness $\delta b \sim 1.5 d$. The $A$--dependence of the cross section is simply
\be
\sigma_{\rm incoh} \sim 3\pi R d  \frac{\sigma_{\rm CEP}^{pp}}{\sigma_{\rm tot}^{nn}}\propto A^{1/3}\;,
\ee
Thus we expect a $\sim A^{1/3}$ scaling, with no additional numerical suppression in the prefactors.

For the coherent case a similar approach can be taken, however instead of \eqref{eq:incohaa} we find
\begin{align}
\sigma_{\rm coh}^{AA} &=4\pi R \,\sigma_{\rm CEP}^{pp}\left(\frac{4\pi}{\sigma_{\rm tot}^{nn} \left \langle q_{\perp}^2 \right \rangle}\right)^2 \frac{1}{(2\pi R d)^{1/2}}\int {\rm d}\Delta \,\frac{P^2 e^{-P} }{\Delta} \;,\\
&=4\pi R \,\sigma_{\rm CEP}^{pp}\left(\frac{4\pi}{\sigma_{\rm tot}^{nn} \left \langle q_{\perp}^2 \right \rangle}\right)^2 \frac{1}{(2\pi R d)^{1/2}}\int {\rm d}P \,\frac{P^2 e^{-P} }{\left|1 - \frac{\Delta}{d}\right|} \;,\\
&\sim \left(\frac{4\pi}{\sigma_{\rm tot}^{nn} \left \langle q_{\perp}^2 \right \rangle}\right)^2 \cdot \sigma_{\rm CEP}^{pp} \cdot A^{1/6}\;,
\end{align}
where in the second line we again use that the dominant part of the integral comes from the $P\sim 1$ region.

\bibliography{references}{}
\bibliographystyle{h-physrev}

\end{document}